\documentclass[%
 aip,
 amsmath,amssymb,
 reprint, onecolumn 
]{revtex4-1}

\usepackage{graphicx}
\usepackage{dcolumn}
\usepackage{bm}

\usepackage[T1]{fontenc}
\usepackage{mathptmx}
\usepackage{etoolbox}
\usepackage{amssymb} 

\usepackage{CJKutf8}

\usepackage{hyperref}
\usepackage{xcolor}
\usepackage{cleveref}
\usepackage{booktabs}
\usepackage{subcaption}
\usepackage{pgfplots}
\usepackage{tikz,pgfplots}
\usetikzlibrary{external}
\pgfplotsset{
    every axis/.append style={
        line width=0.6pt,
    },
}

\usepackage{multirow}
\usepackage{caption}
\usepackage{threeparttable}
\usepackage{tabularx}
\usepackage{array} 
\usepackage[normalem]{ulem}
\usepgfplotslibrary{fillbetween}
\pgfplotsset{compat=1.9}
\usetikzlibrary{arrows.meta}
\newcolumntype{L}{>{\raggedright\arraybackslash}X} 
\newcolumntype{C}{>{\centering\arraybackslash}X}   

\pgfplotsset{
    mygridstyle/.style={
        grid style=solid, 
        line width=0.35pt, 
        color=gray!50, 
    },
}
\usepackage{adjustbox}
\usepackage{footmisc}
\usepackage{threeparttable}

\definecolor{hcorange}{RGB}{255,127,80}
\definecolor{hcblue}{RGB}{30,144,255}
\definecolor{hcgreen}{RGB}{229,242,229}
\definecolor{hcpurple}{RGB}{229,229,255}
\definecolor{hcpurple1}{RGB}{155,110,188}
\definecolor{hcred}{RGB}{130,118,213}
\definecolor{hc03}{RGB}{255,99,72}
\definecolor{hc15}{RGB}{47,53,66}

\crefname{equation}{Eq.}{Eqs.}
\Crefname{equation}{Eq.}{Eqs.}
\crefname{figure}{Fig.}{Figs.}
\Crefname{figure}{Fig.}{Figs.}
\crefname{algorithm}{Algorithm}{Algorithms}
\Crefname{algorithm}{Algorithm}{Algorithms}
\crefname{section}{Section}{Sections}
\Crefname{section}{Section}{Sections}
\crefname{table}{Table.}{Tables.}
\Crefname{table}{Table.}{Tables.}

\makeatletter
\def\@email#1#2{%
 \endgroup
 \patchcmd{\titleblock@produce}
  {\frontmatter@RRAPformat}
  {\frontmatter@RRAPformat{\produce@RRAP{*#1\href{mailto:#2}{#2}}}\frontmatter@RRAPformat}
  {}{}
}%
\makeatother

\begin{document}
\preprint{AIP/123-QED}

\title[Robust DRL for enhancing flow control]{Robust and Adaptive Deep Reinforcement Learning for Enhancing Flow Control around a Square Cylinder with Varying Reynolds Numbers}

\author{Wang Jia (\begin{CJK*}{UTF8}{gbsn}贾旺\end{CJK*})}
\author{Hang Xu (\begin{CJK*}{UTF8}{gbsn}徐航\end{CJK*})}
 \email{hangxu@sjtu.edu.cn}
\affiliation{State Key Lab of Ocean Engineering, School of Naval Architecture, Ocean and Civil Engineering, Shanghai Jiao Tong University, Shanghai, 200240, China}

\date{\today}

\begin{abstract}

The present study applies a Deep Reinforcement Learning (DRL) algorithm to Active Flow Control (AFC) of a two-dimensional flow around a confined square cylinder. Specifically, the Soft Actor-Critic (SAC) algorithm is employed to modulate the flow of a pair of synthetic jets placed on the upper and lower surfaces of the confined squared cylinder in flow configurations characterized by $Re$ of 100, 200, 300, and 400. The investigation starts with an analysis of the baseline flow in the absence of active control. 
It is observed that at $Re = 100$ and $Re = 200$, the vortex shedding exhibits mono-frequency characteristics. Conversely, at $Re = 300$ and $Re = 400$, the vortex shedding is dominated by multiple frequencies, which is indicative of more complex flow features. With the application of the SAC algorithm, we demonstrate the capability of DRL-based control in effectively suppressing vortex shedding, while significantly diminishing drag and fluctuations in lift. Quantitatively, the data-driven active control strategy results in a drag reduction of approximately 14.4\%, 26.4\%, 38.9\%, and 47.0\% for $Re = 100$, 200, 300, and 400, respectively. To understand the underlying control mechanism, we also present detailed flow field comparisons, which showcase the adaptability of DRL in devising distinct control strategies tailored to the dynamic conditions at varying $Re$. These findings substantiate the proficiency of DRL in controlling chaotic, multi-frequency dominated vortex shedding phenomena, underscoring the robustness of DRL in complex AFC problems. 

\end{abstract}

\maketitle

\section{Introduction}\label{sec:introduction}
The examination of fluid dynamics around square cylinders is of paramount importance within both scholarly inquiry and practical engineering realms, encompassing applications such as maritime structures, skyscrapers, towers, bridge piers, and heat exchangers.\cite{bhatt2018vibrations,bai2018dependence,Flowpast2010,Subhankar}
The vortex shedding phenomenon of square cylinders, that is, the alternating shedding vortices formed behind the square cylinder, will generate periodic lateral forces, which may cause structural fatigue, resonance, noise, reduced performance, and safety risks.\cite{NumericalZhao}
In view of this, mitigating the adverse effects of fluid forces around square cylinders has become a critical research effort.
Flow control technology changes the flow pattern of fluid and is widely used in marine, construction, aviation, automobile, energy and other fields.\cite{rastan2019controlled} It can reduce drag, improve aerodynamic performance, increase lift or load-bearing capacity, reduce vibration and noise, and control heat and mass transfer, which is of great significance to improving efficiency, performance and sustainable development. Active flow control technology, with its ability to sense and accurately regulate fluid flow in real time, demonstrates significant performance advantages over traditional passive control methods.\cite{ActiveGao,rastan2019controlled}

While active flow control technology has advanced significantly, its widespread adoption has been limited by high cost and technical complexity, reliance on high-performance computing resources, the need for a deep understanding of fluid dynamics during design and optimization, and the environment challenges with variability in operating conditions.\cite{rabault2020deep,ren2020active} 
The rapid evolution of machine learning in recent years has not only achieved significant advancements within the domain of computer science but has also exerted profound impacts on specialized fields such as fluid dynamics and aerodynamics.
In particular, the swift progression of machine learning has introduced new opportunities for research in active flow control, catalyzing progress within this sector.\cite{annurevfluid}
The spatiotemporal evolution of flow is predominantly governed by the nonlinear Navier-Stokes equations, which inherently encapsulate complex characteristics such as high-dimensionality, multifrequency, multimodality, and multiscale phenomena. These complexities substantially elevate the challenges and intricacies associated with flow control, rendering traditional analytical and numerical approaches computationally intensive and sometimes intractable.\cite{rabault2020deep,Vignon2023,XieFangfang2023,ren2020active} 
Under such circumstances, Deep Reinforcement Learning (DRL) offers a promising new solution to the flow control problem. Previous studies have demonstrated the potential of DRL in Active Flow Control (AFC) applications. For example, DRL has been successfully used to control flow separation around airfoils, reduce drag on bluff bodies, and optimize flow control actuators.\cite{RabaultKuhnle2023,GARNIER2021104973,fluids7020062} These studies demonstrate that DRL can effectively learn complex control strategies and adapt to different flow conditions, outperforming traditional control methods in terms of performance and efficiency.\cite{Popat,viquerat2022review} Overall, the clever application of DRL in AFC demonstrates its potential to solve complex decision-making problems.

Deep learning (DL) has demonstrated remarkable proficiency in extracting nonlinear features from complex systems, such as chaotic systems.\cite{annurevfluid,bruntonClosed} DL can learn and represent highly abstract features, thereby capturing the nonlinear patterns and dynamic changes inherent in complex systems.\cite{mahesh2020machine,sarker2021machine,burrell2016machine} 
This feature enables DL to effectively extract features of complex systems.\cite{sarker2021machine,burrell2016machine}
Reinforcement learning (RL) is a key subfield of machine learning that has attracted significant attention in recent years, particularly after the integration of deep neural networks to create the Deep Q-Network (DQN) artificial intelligence agent. The DQN, by learning from high-dimensional inputs, has demonstrated remarkable performance in various complex tasks, marking an important milestone.\cite{mnihHumanlevelControlDeep2015}
The triumph of AlphaGo over the world Go champion in 2016 further highlighted RL's adeptness at complex decision-making and strategic planning.\cite{silver2017mastering} These developments have significantly propelled the interest and perceived impact of DRL within the AI domain. 
The synergistic ability of DL combined with RL can solve decision-making problems that were previously considered intractable, especially in non-linear and high-dimensional fusion scenarios.\cite{kaelbling1996reinforcement,8103164} 

DRL provides an innovative approach to optimizing fluid flow control strategies in various areas, including wall turbulence control\cite{guastoni2023discovering}, channel flow\cite{guastoni2023deep}, convection control\cite{WANG2023123655}, heat transfer\cite{HACHEM2021110317,WANG2023123655}, combustion turbulence\cite{CHENG2018303,ZhanXu2022}, and bluff bodies\cite{Amico2022}. The first application of neural networks for active turbulence control on walls was introduced by \citeauthor{Lee}\cite{Lee}. They used blowing/suction devices to lower shear stress in turbulent channel flows. Their approach, leveraging Artificial Neural Networks (ANN) for closed-loop control, resulted in a notable drag reduction of up to 20\%. 
\citeauthor{lee2023turbulence}\cite{lee2023turbulence} used DRL for turbulence control, achieving a 20\% reduction in drag. This study demonstrates the efficacy of DRL in turbulence control and provides a physical interpretation of its mechanisms.
\citeauthor{guastoni2023deep,guastoni2023discovering}\cite{guastoni2023deep,guastoni2023discovering} employed DRL algorithms to reduce the drag of turbulent flows confined within channels, achieving drag reductions of 43\% in minimal channels and 30\% in larger channels.
\citeauthor{Gerben}\cite{Gerben} successfully utilized small temperature fluctuations to reduce heat transfer in a two-dimensional Rayleigh-Bénard system, applying advanced DRL algorithms. 
\citeauthor{WANG2023123655}\cite{WANG2023123655} harnessed a DQN in synergy with three intricate techniques to elucidate the core mechanisms of heat and mass transfer in the ambit of closed-loop active control interactions.

DRL has also been successfully applied to the study of flow control strategies in confined cylindrical flow and blunt body flow scenarios. \citeauthor{rabaultArtificial}\cite{rabaultArtificial} pioneered the use of DRL for AFC, achieving an 8\% reduction in cylinder resistance in a 2D simulation at $Re$ of 100. 
This work is the first to release open-source code that combines DRL with CFD coupling. In his review paper, he described how to use the class functions in the \texttt{Tensorforce} library to embed your own CFD environment into the DRL algorithm.\cite{rabault2020deep}
\citeauthor{liReinforcementlearning}\cite{liReinforcementlearning} successfully controlled the vortex shedding phenomenon behind a confined cylinder by embedding the physical information of hydrodynamics into the design of the reward function. Not only embedded physical information in the DRL control strategy but also validated the robustness of DRL-based control methods for various blockage ratios of confined cylinders.
\citeauthor{parisRobustFlowControl2021a}\cite{parisRobustFlowControl2021a} investigated the introduction of S-PPO-CMA optimization for sensor placement. At a $Re$ of 120, they utilized this algorithm to train a control strategy that achieved an 18.4\% reduction in drag.
\citeauthor{wangDRLinFluids}\cite{wangDRLinFluids} introduced DRLinFluids, a Python-based platform for flow control, achieving drag reductions of approximately 8\% and 13.6\% for circular and square cylinders, respectively, at $Re = 100$.
 
\citeauthor{rabaultArtificial,tangRobustActiveFlow2020,heess2017emergence,renApplying,jia2024optimal}\cite{rabaultArtificial,tangRobustActiveFlow2020,heess2017emergence,renApplying,jia2024optimal} have applied the Proximal Policy Optimization (PPO) algorithm to AFC in the flow around a cylinder, and the research results consistently demonstrate a reduction in drag of approximately 8\% at $Re = 100$.
\citeauthor{tangRobustActiveFlow2020}\cite{tangRobustActiveFlow2020} utilized a flow control strategy based on DRL to actively reduce drag. They achieved approximately 21.6\%, 32.7\%, and 38.7\% reduction in drag at \(Re = 200,300,400\), respectively. 
\citeauthor{renApplying}\cite{renApplying} explored the efficacy of DRL in managing weak turbulence at \(Re = 1000\), finding that DRL agents could identify strategies to reduce drag by up to 30\%. 
The aforementioned studies comprehensively examined strategies for AFC using DRL, with a focus on flow around a confined cylinder as the test case. These investigations spanned multiple dimensions, including the robustness of the $Re$ ranging from 100 to 1000, the exploration of probe positions and quantities, the incorporation of physical information into the reward function, and the enhancement of the PPO algorithm. The study on flow around a confined cylinder offers a comprehensive observation and analysis of DRL's application in AFC, providing valuable insights for this field. 

Square cylinder flow and circular cylinder flow exhibit significant differences in physical phenomena and flow characteristics, primarily due to their geometric shapes and the fundamental principles of fluid dynamics.\cite{bhatt2018vibrations,bai2018dependence,Flowpast2010} Square cylinders are prone to flow instabilities at lower Reynolds numbers, such as the von Kármán vortex street, with separation points fixed at sharp edges, while circular cylinders see separation points that move rearward with increasing Reynolds number.\cite{bhatt2018vibrations,bai2018dependence,Flowpast2010,Subhankar} Additionally, vortex shedding around square cylinders occurs earlier and is more complex compared to the regular vortex patterns seen with circular cylinders.\cite{bhatt2018vibrations,Flowpast2010,Subhankar,Sohankar,Arun,bai2018dependence,sen2009steady,tritton1959experiments,SAHA200354} 
The influence of Reynolds number on drag and lift is more pronounced in square cylinder flow.\cite{bhatt2018vibrations,bai2018dependence,Flowpast2010,Subhankar,park1998numerical}
Understanding these differences is crucial for the design and optimization of fluid control systems.


In recent years, there has been some research applying DRL algorithms to AFC control strategies within the context of flow around square cylinders.
In the study conducted by \citeauthor{wangDRLinFluids}\cite{wangDRLinFluids}, the investigation was exclusively focused on a $Re$ of 100, where the drag reduction rate was observed to be 13.7\%, and the wake vortices continued to exhibit an oscillating state.
\citeauthor{chen2023deep}\cite{chen2023deep} used a DRL-based AFC method to reduce the vortex-induced vibration of a square cylinder when the $Re = 100$ , and placed the injection actuators at the front, middle, and rear corners of the side walls of the square cylinder. The control results show that when the jet actuator is located at the rear corner, the control effect is faster and more obvious than at the front corner and middle position. The control results show that when the jet actuator is located at the rear corner, the drag reduction is the most effective with the least actuator energy consumption. \citeauthor{yan2023stabilizing}\cite{yan2023stabilizing} studied the effect of synthetic jet placement on drag and lift around a square cylinder using Soft Actor-Critic (SAC) algorithm. They found jets at the front edge more effective than those placed at the rear, reducing drag coefficient by up to 44.4\%, 60.6\%, and 57.8\% at $Re$ of 500, 1000, and 2000, respectively.
In another work by \citeauthor{yan2024aero}\cite{yan2024aero}, they investigated the effect of using a combination of eight jets located at four corners of rectangular cylinders at $Re$ of 1000. With four or eight jets activated, they were able to reduce drag by 63.2\% and 77.1\% respectively on a square cylinder.

Despite the significant achievements of AFC technology based on DRL in reducing the drag and mitigating lift fluctuations around square cylinders, a primary issue remains: the instability of the wake flow and the phenomenon of vortex shedding have not been effectively controlled. In the circular cylinder case, vortex shedding can be almost fully suppressed and only very weak jet flow is required to maintain the controlled state\cite{wangDRLinFluids}. Such a fully controlled vortex shedding pattern with minimal control energy has not been observed in existing literature for the square cylinder test case, even at a relatively low $Re$ of 100. To the best of our knowledge, only one test case in the work by \citeauthor{chen2023deep}\cite{chen2023deep} showed signs of stabilized wake flow at $Re = 100$.
The current state of the art is capable of sustaining the drag reduction throughout the control period, but with substantial vortex shedding in the downstream, which could induce adverse vibrations on the square cylinder itself or any downstream structures. Due to the instability of the wake flow and the persistence of vortex shedding, delving into the physical mechanisms behind DRL-based AFC strategies also remains a challenge. Therefore, a detailed investigation into the control effects of DRL-based AFC technology in managing the flow around square cylinders is particularly crucial to achieve a more stable wake flow and to deepen the understanding of its physical mechanisms.

In this study, we aim to explore the potential of DRL-based AFC technology in controlling the flow around square cylinders to suppress vortex shedding with minimal actuator excitation energy. To investigate the control mechanisms devised by DRL at various flow conditions, we examine flows with $Re$ from 100 to 400. The development of an efficient, energy-saving, stable, and robust active flow control strategy is intended to deepen the understanding of the control mechanisms behind DRL-based AFC technology and promote its application in engineering practices, thereby advancing the technology both theoretically and practically. In \cref{sec:Problem setup and methodology}, the problem description is presented, followed by an in-depth overview of DRL and its formulation in addressing the AFC challenge. 
The main results and discussions are developed in \cref{sec:Results and discussion}.
Finally, \cref{sec:Conclusions} concludes the paper by summarizing the key findings, discussing their implications, and suggesting potential directions for future research.

\section{Problem setup and methodology}\label{sec:Problem setup and methodology}
\subsection{Numerical simulation setup}
\begin{figure*}[ht]
    \centering
    \begin{subfigure}{0.47\textwidth}
    \includegraphics[width=\textwidth]{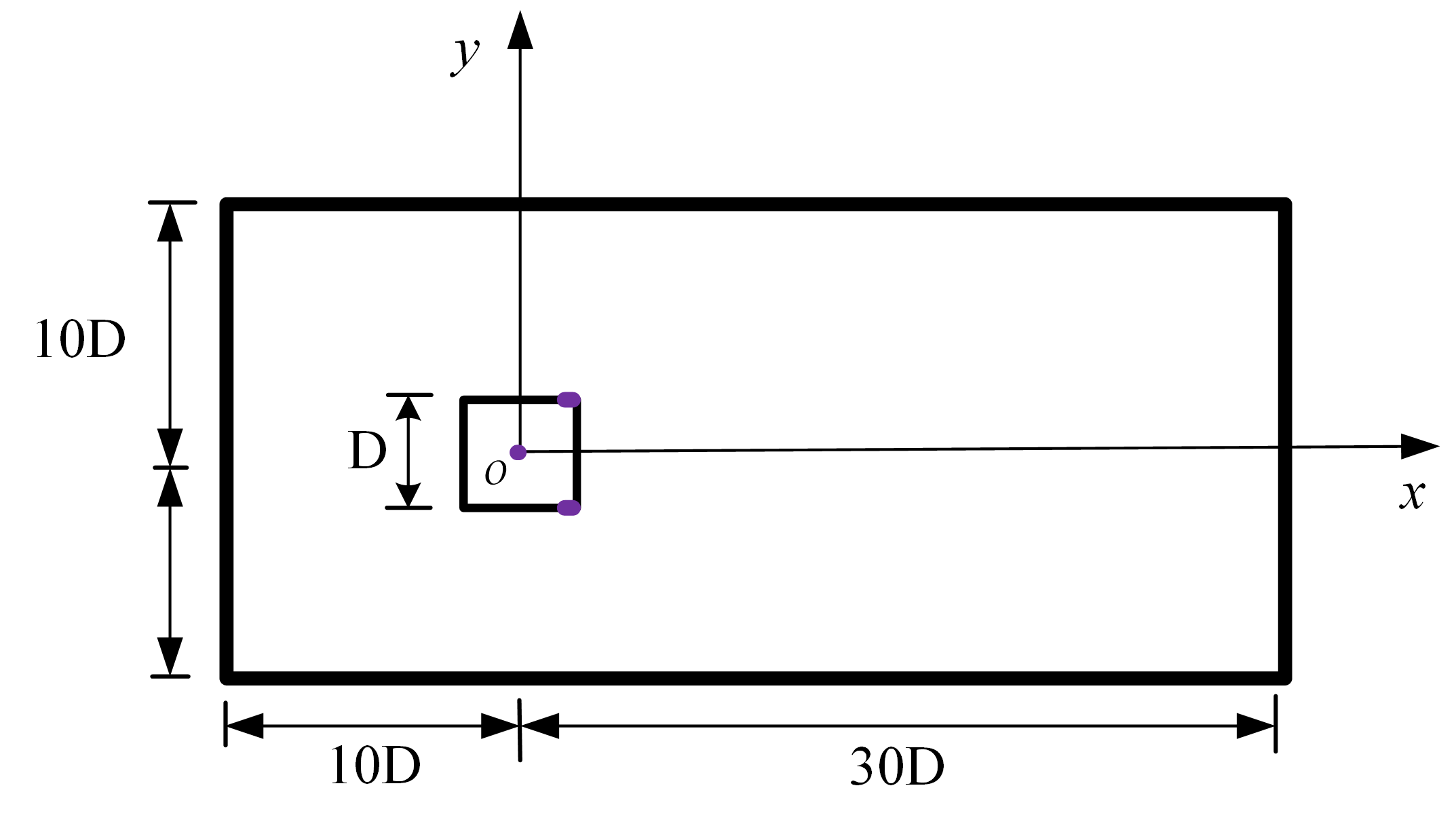}
    \caption{}
    \label{fig:square01}
    \end{subfigure}
    \hspace{5mm} 
    \begin{subfigure}{0.26\textwidth}
    \includegraphics[width=\textwidth]{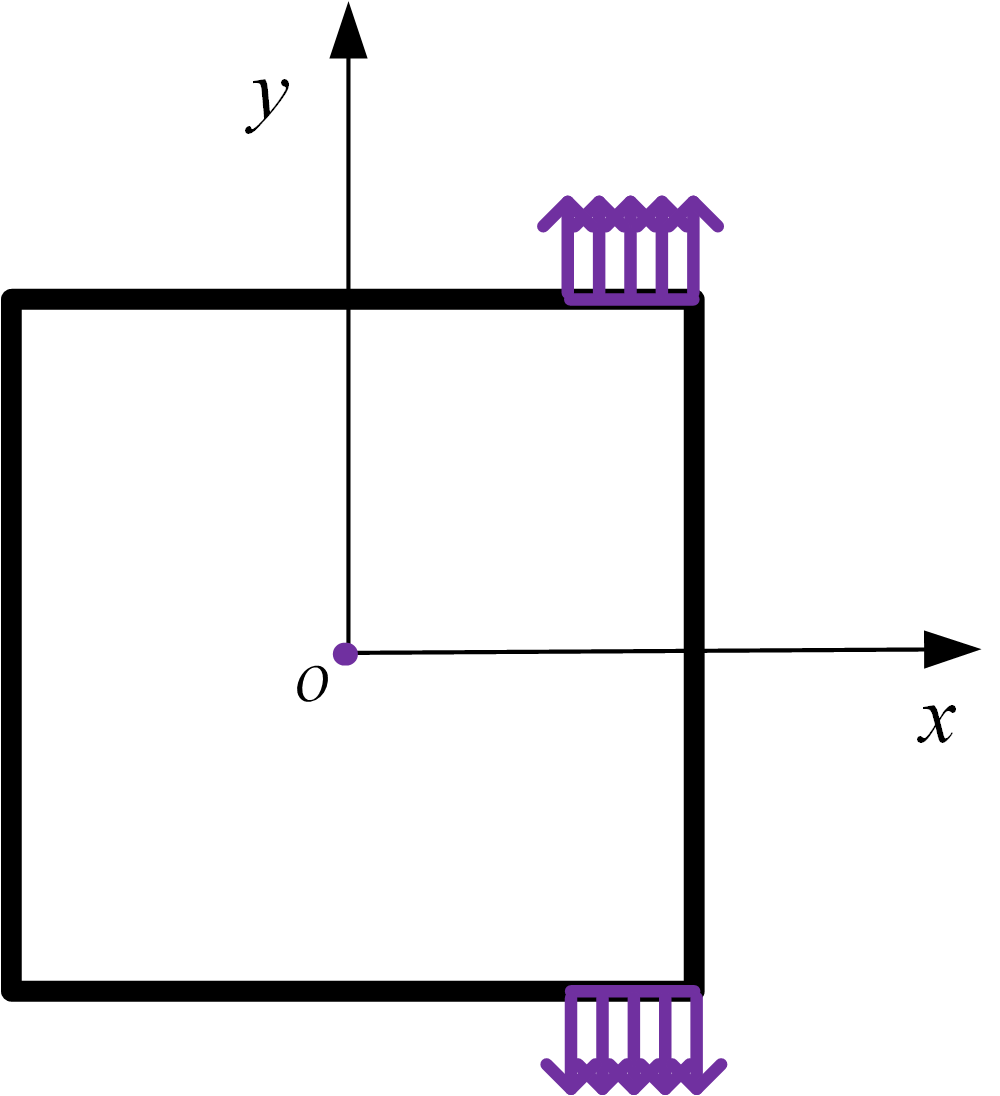}
    \caption{}
    \label{fig:square02}
    \end{subfigure}
    \caption{Description of the 2D square cylinder numerical setup: (a) The computational domain has a length of $40D$ and a width of $20D$, with the square cylinder located at the center of the domain. Each side of the square cylinder is of length $D$. A uniform upstream velocity flows in the positive $x$-direction past the square cylinder towards the downstream. (b) This figure only shows the starting position of the jets on the square cylinder, with the scale ratios not reflecting actual proportions.}
    \label{fig:square}
\end{figure*}

\paragraph{Model configuration}
The configuration studied in this paper is the two-dimensional (2D) flow around a square cylinder. The side length of the square cylinder is $D$.
\Cref{fig:square01} illustrates the computational domain and coordinate system of the numerical model. The origin of the Cartesian coordinate system is established at the center of the square cylinder. The square cylinder is immersed in a rectangular domain of dimensions $40D$ (along the $x$-axis) and $20D$ (along the $y$-axis). \cref{fig:square02} is a detailed view of the square cylinder, showing the position of a symmetrical pair of jets.

\paragraph{Governing equations}

To establish the physical model for numerical simulations, it is assumed that the flow is viscous and incompressible. Flow simulations are executed by solving the two-dimensional incompressible Navier-Stokes equations within the computational domain.
The governing equations can be represented as follows:
\begin{equation}\label{eq:ns1}
\frac{\partial \boldsymbol{u}}{\partial t}+\boldsymbol{u} \cdot(\nabla \boldsymbol{u})=-\nabla p+\frac{1}{Re} \Delta \boldsymbol{u},
\end{equation}
\begin{equation}\label{eq:ns2}
\nabla \cdot \boldsymbol{u}=0.
\end{equation}

Here, $\boldsymbol{u} = (U, V)$ represents the velocity vector, where $U$ and $V$ are the velocity components along the $x$- and $y$-axes, respectively. $t$ denotes the dimensionless time unit. $p$ is the thermodynamic pressure. $Re = \frac{\overline{U}D}{\nu}$ is the $Re$, where $D$ is the the side length of the square,$\nu$ denotes the kinematic viscosity of the fluid. and $\overline{U}$ is the mean velocity at the inlet.

\paragraph{Boundary conditions}
The left boundary of the computational domain is defined as the velocity inlet $\Gamma_\text{in}$, with the boundary conditions, $U=U_0$, $V=0$. 
In this study, a uniform flow with a velocity of $U_0=2$ is imposed.
The right boundary is defined as the pressure outlet $\Gamma_\text{out}$, with the boundary condition, the mean static pressure is set to zero. 
The spanwise boundaries are defined as symmetric boundary conditions $\Gamma_\text{s}$, with the boundary condition, the velocity component normal to each variable is zero. 
The positions of the two synthetic jet actuators near the trailing edge of the square cylinder are defined as the jet boundaries $\Gamma_{i}\, (i = 1,2)$. The boundary conditions impose a uniform velocity distribution along the normal direction. 
The width of the synthetic jets velocity distribution is $0.04D$. The jet velocities of the synthetic jets can be positive or negative, corresponding to blowing or suction, respectively. Maintaining the balance of the net mass flow rate for all jets ensures a zero net flow rate for the system, i.e., $V_{\Gamma_1} = -V_{\Gamma_2}$.
Apart from the jets, the rest of the square cylinder adopts no-slip solid wall boundary conditions $\Gamma_\text{w}$, with the boundary conditions, $U=0$, $V=0$. 

\paragraph{Solver details}
In this study, the incompressible flow solver is based on the open-source CFD package \texttt{OpenFOAM}, as described by Jasak \textit{et al.}\cite{jasakOpenFOAMOpenSource2009,jasakOpenFOAMLibraryComplex2013} \texttt{OpenFOAM} is a widely used and validated CFD software package that provides robust numerical algorithms for solving the Navier-Stokes equations. The solver in \texttt{OpenFOAM} utilizes the finite volume method to discretize the computational domain into a mesh made up of control volumes.
To ensure numerical stability, the time step is chosen as $\Delta t$ = 0.0005. 

\begin{figure*}[ht]
    \centering
    \begin{subfigure}{0.4\textwidth}
    \includegraphics[width=\textwidth]{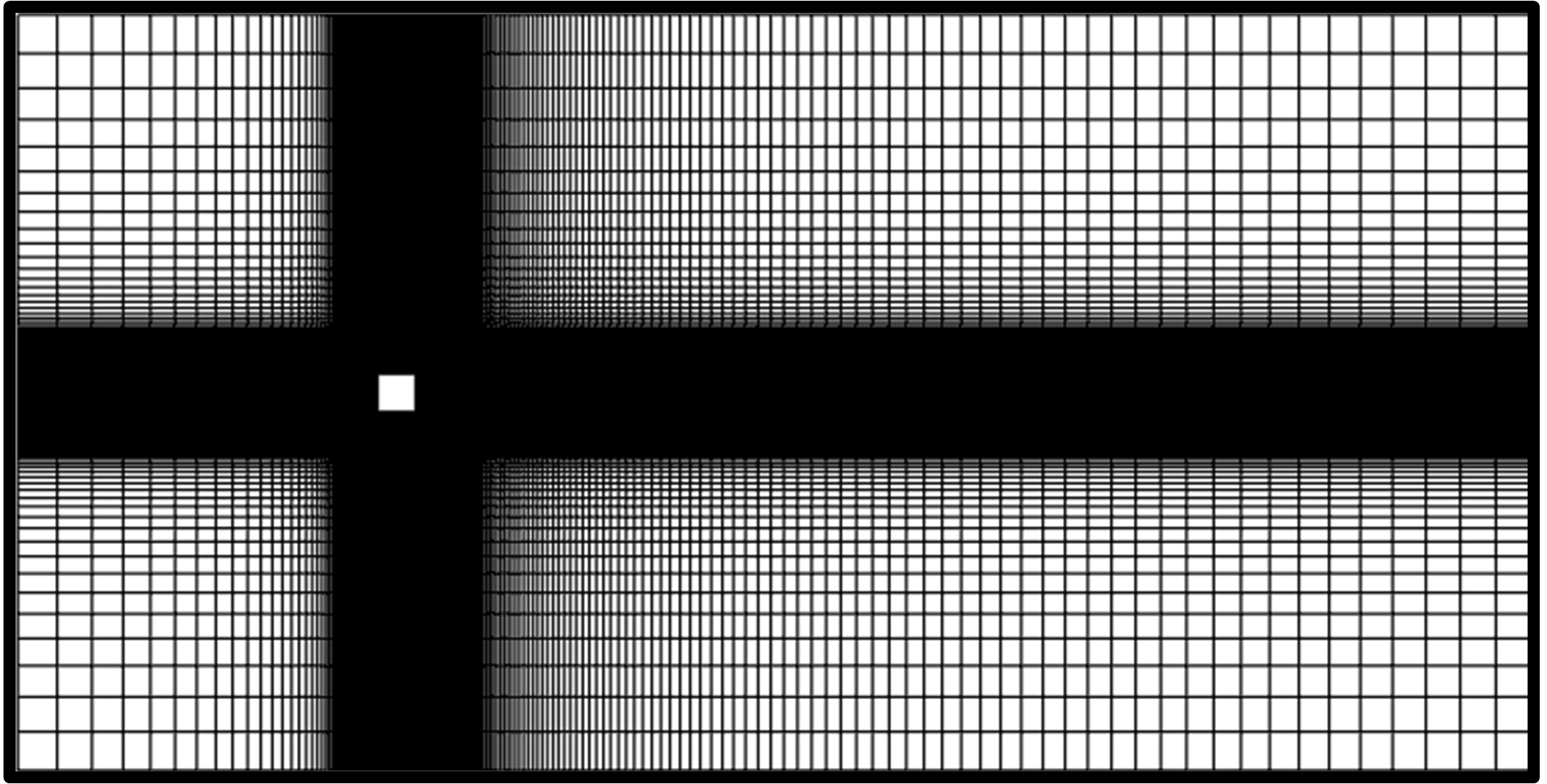}
    \caption{}
    \label{fig:mesh01}
    \end{subfigure}
    \hspace{6mm} %
    \begin{subfigure}{0.22\textwidth}
    \includegraphics[width=\textwidth]{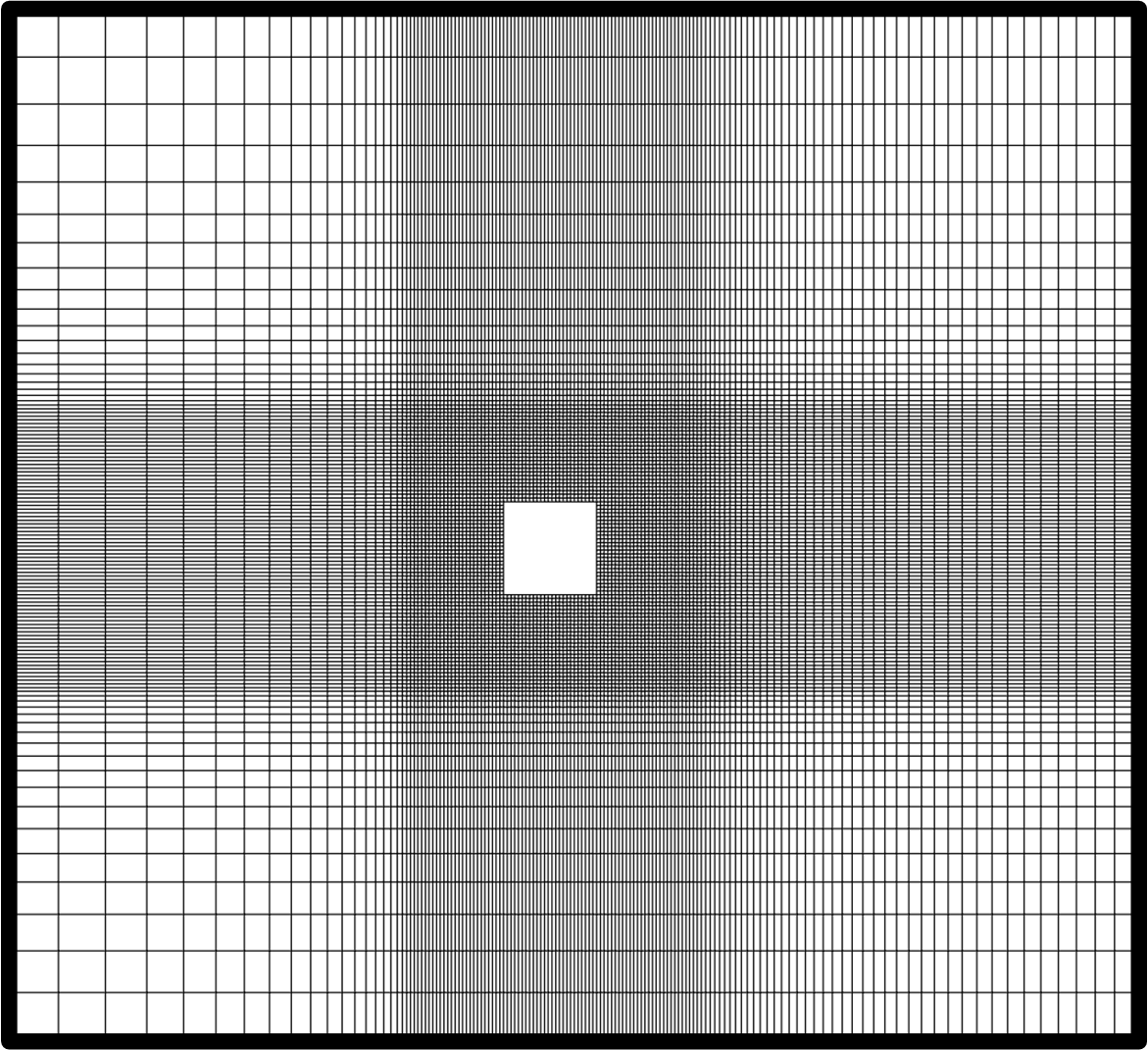}
    \caption{}
    \label{fig:mesh02}
    \end{subfigure}
     \hspace{6mm} %
    \begin{subfigure}{0.4\textwidth}
    \includegraphics[width=\textwidth]{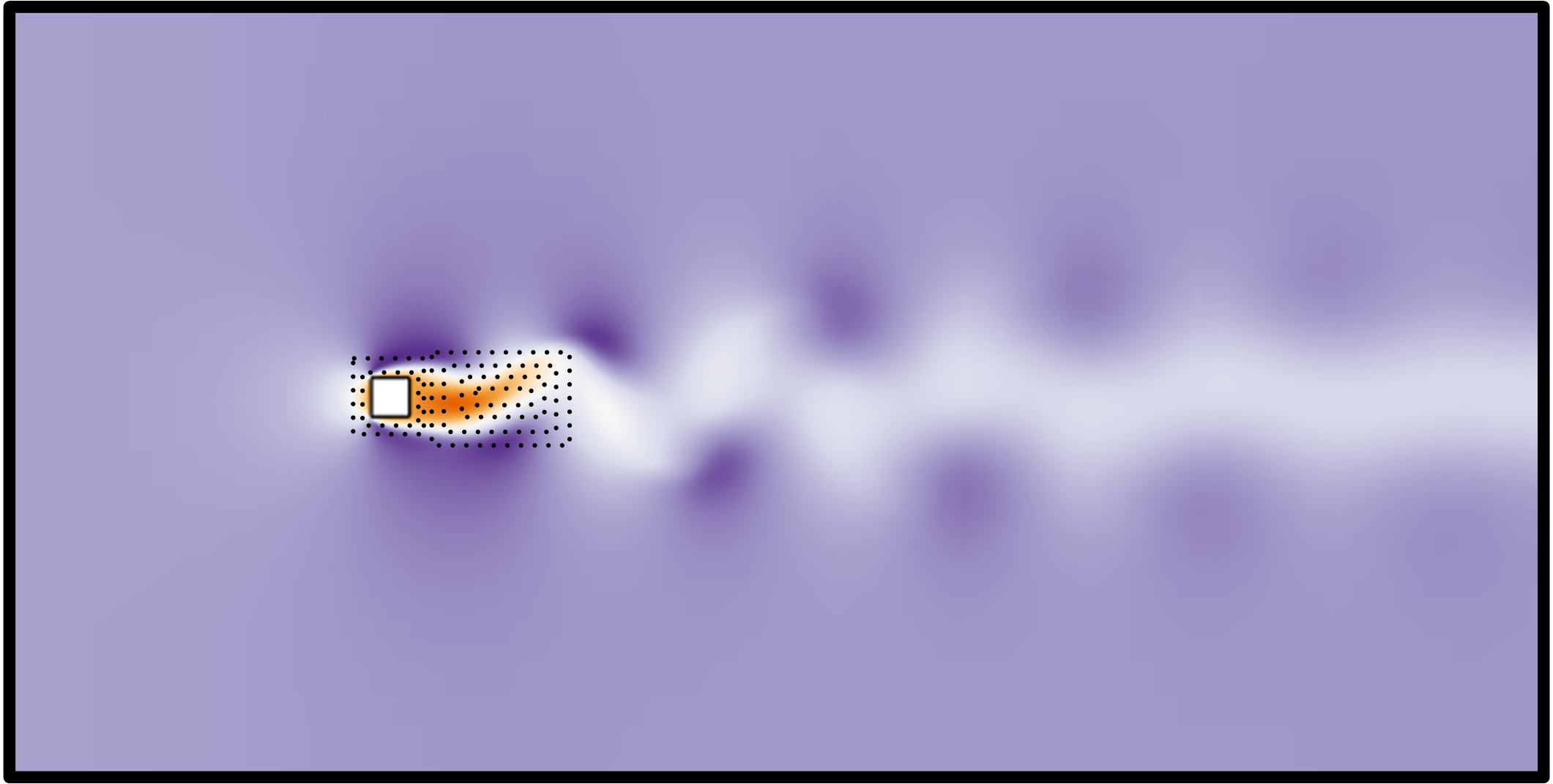}
    \caption{}
    \label{fig:mesh03}
    \end{subfigure}
    \hspace{5mm} %
    \begin{subfigure}{0.22\textwidth}
    \includegraphics[width=\textwidth]{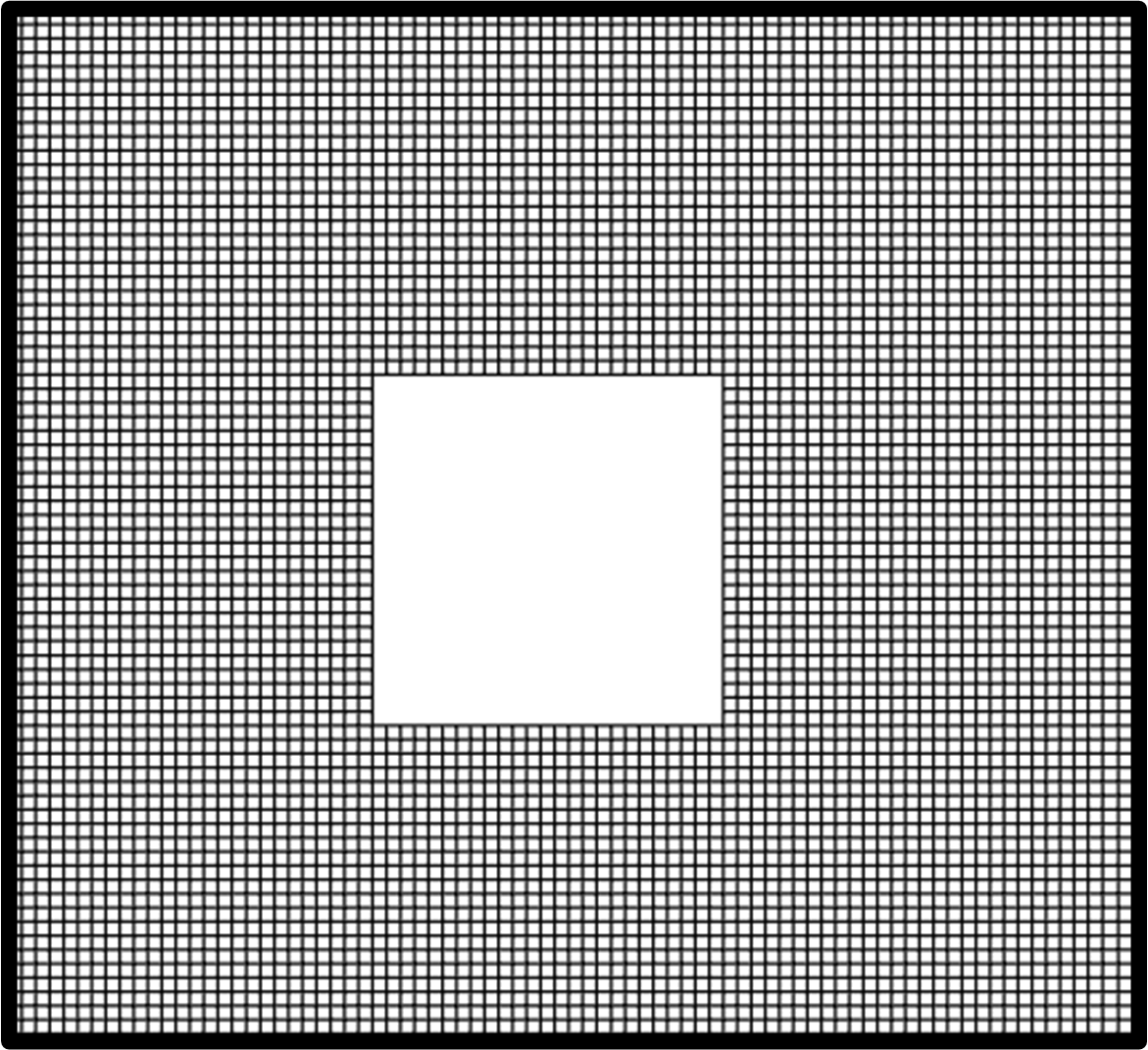}
    \caption{}
    \label{fig:mesh04}
    \end{subfigure}
    \caption{Discretization of the computational domain. (a) Meshing of the global computational domain; (b) Mesh division near the square cylinder; (c) The velocity field around the square cylinder after the initiation of flow without flow control. The black dots represent the positions of 201 velocity probes; (d) More detailed mesh division around the square cylinder.}
    \label{fig:mesh}
\end{figure*}

\paragraph{Grid system}
The computational domain is discretized using structured meshes and is refined around the surface of the square cylinder and in the wake flow region downstream of the cylinder, as shown in \cref{fig:mesh}. 
The computational domain is discretized into 23,264 grid elements using structured quadrilateral cells.
\cref{fig:mesh01} presents the global discretization of the computational domain, while \cref{fig:mesh02} depicts a detailed treatment around the square cylinder to facilitate consideration of the impact of on flow simulation. 
\cref{fig:mesh03} displays the velocity field around the square cylinder in the absence of flow control, primarily to show the placement of 201 probes. \cref{fig:mesh04} details the mesh around the square cylinder with a closer view.

\paragraph{Quantities of interest}
The lift coefficient ($C_L$) and drag coefficient ($C_D$) are defined as
\begin{align}
     C_L = \frac{F_L}{0.5\rho \overline{U}^2D}, \quad  C_D = \frac{F_D}{0.5\rho \overline{U}^2D}.
\end{align}
Here, $F_L$ and $F_D$ represent the lift and drag forces integrated on the surface of the cylinder, respectively, and $\rho$ is the fluid density.
The Strouhal number $(St)$ is used to describe the characteristic frequency of oscillatory flow phenomena and is defined as follows:
\begin{equation}
St = \frac{f \cdot D}{U}.
\end{equation}
Where $f_s$ is the shedding frequency calculated based on the periodic evolution of the $C_L$.
$D$ is the characteristic length, which is the side length of the square cylinder, and $\overline{U}$ is the mean velocity of the upstream flow.

\paragraph{Grid independence}
To investigate grid convergence and validate the numerical methods, the flow under $Re = 100$ was simulated to compute the desired quantities of interest. These results were then compared with the computations carried out by other published paper.

\begin{table*}[htbp]
\centering
\begin{threeparttable}
\caption{Mesh convergence and flow parameters for the 2D flow around a square cylinder at $Re = 100$.}
\label{tab:meshconvergence}
\vspace{-\baselineskip}
\begin{tabularx}{0.93\textwidth}{
  >{\centering\arraybackslash}X
  >{\centering\arraybackslash}X
  >{\centering\arraybackslash}X
  >{\centering\arraybackslash}X
  >{\centering\arraybackslash}X
  >{\centering\arraybackslash}X
  >{\centering\arraybackslash}X
}
\toprule
Case & Mesh resolution & \( \Delta t \) &  \(\overline{C}_D\) & \(C_{D,\max}\) & \(C_{L,\max}\) & \(St\) \\
\midrule
\citeauthor{wangDRLinFluids}\cite{wangDRLinFluids} & 23 125 & 0.0005 & 1.549 & - & - & - \\ 
Coarse & 8465 & 0.0005 & 1.532 & 1.562 & 0.326 & 0.145 \\
Main & 23 264 & 0.0005 & 1.548 & 1.559 & 0.321 & 0.142 \\
Fine & 35 179 & 0.0005 & 1.549 & 1.561 & 0.320 & 0.141 \\
\bottomrule
\end{tabularx}
\begin{tablenotes}
\item Note: \( \Delta t \) represents the time step size for CFD.
\end{tablenotes}
\end{threeparttable}
\end{table*}

\cref{tab:meshconvergence} lists the simulation results using meshes of three different resolutions and compares them with the results calculated by Wang et al. $C_{D,\max}$ and $C_{L,\max}$ correspond to the maximum values of $C_D$ and $C_L$, respectively. The $C_D$ and $C_L$ calculated with the coarse mesh deviate significantly from the results obtained by \citeauthor{wangDRLinFluids}\cite{wangDRLinFluids} , rendering the precision unreliable. The fine mesh entails a substantial number of grid points, which is disadvantageous for subsequent training processes. This study aims to select discretization results with as few grid points as possible while ensuring accuracy, as this greatly influences the speed of training. The resolution of the main mesh used in this work is sufficiently high to achieve good agreement with the benchmark data.

\subsection{Deep Reinforcement learning}\label{sec:DRL algorithm} 

\begin{figure*}[ht]
    \centering
    \begin{subfigure}{0.355\textwidth}
    \includegraphics[width=\textwidth]{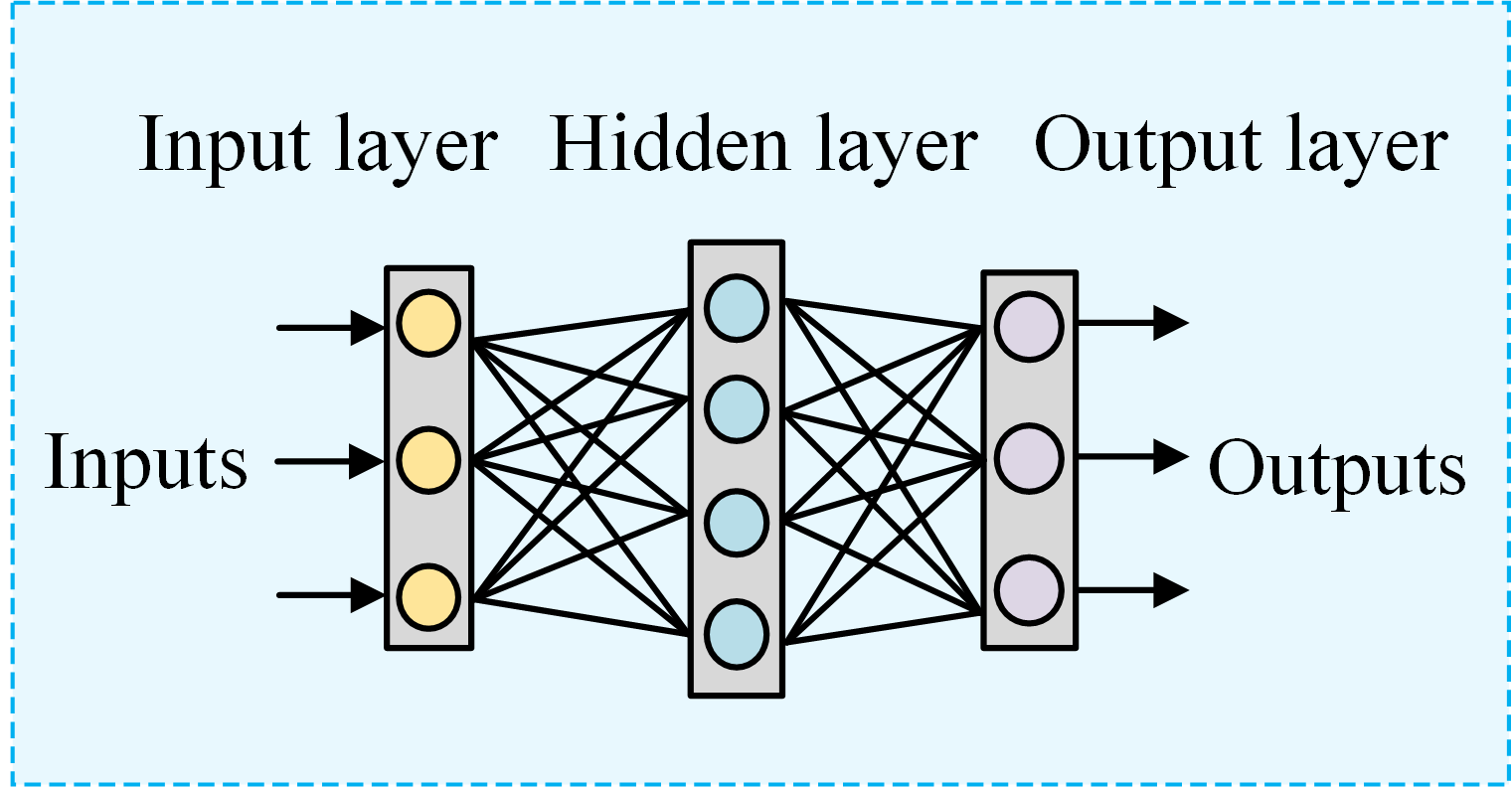}
    \caption{RL}
    \label{fig:ML01}
    \end{subfigure}
    \begin{subfigure}{0.27\textwidth}
    \includegraphics[width=\textwidth]{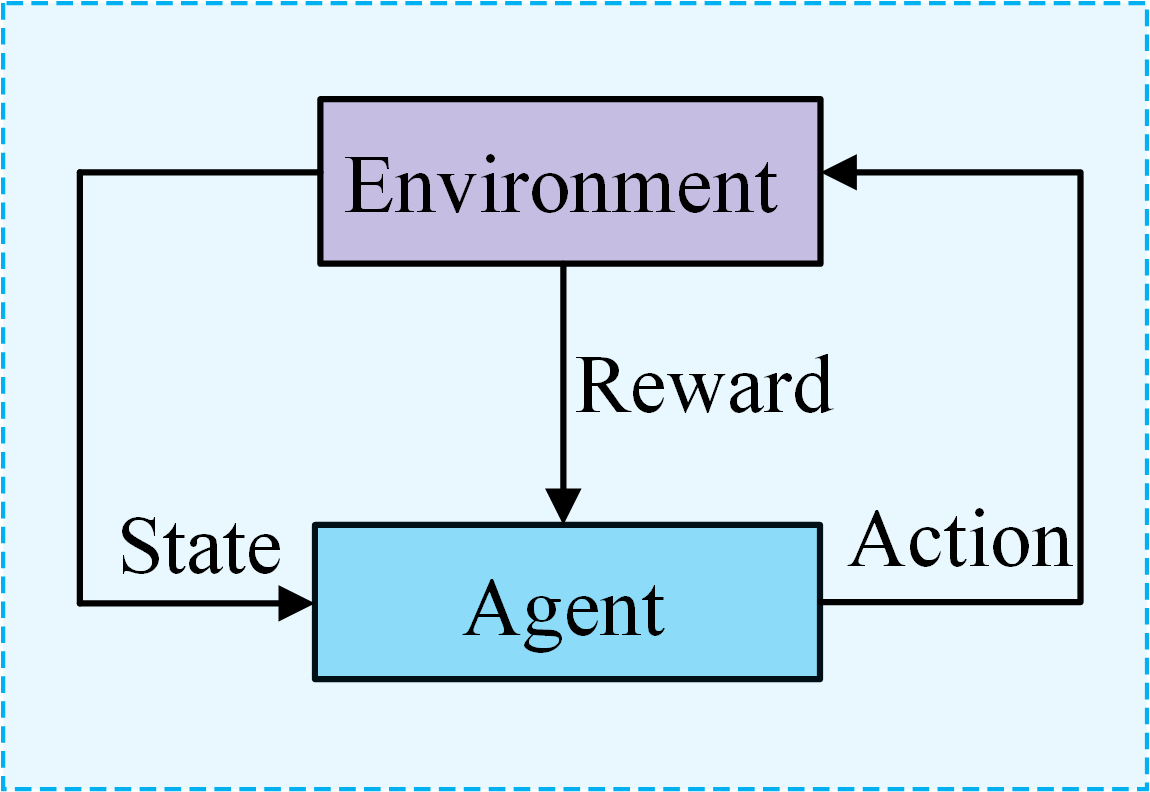}
    \caption{DNN}
    \label{fig:ML02}
    \end{subfigure}
    \begin{subfigure}{0.3\textwidth}
    \includegraphics[width=\textwidth]{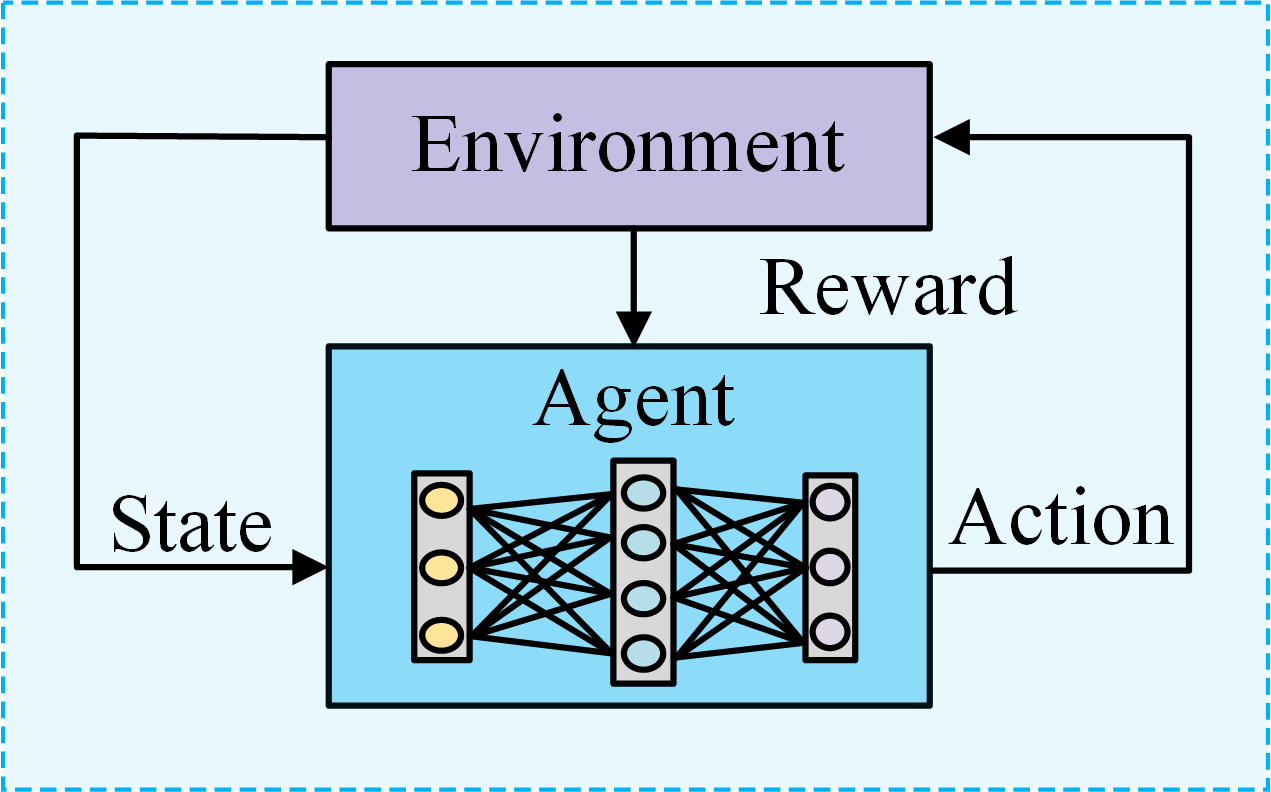}
    \caption{DRL}
    \label{fig:ML03}
    \end{subfigure}
    \caption{Illustration of deep learning, RL, and DRL. (a) In RL, the agent interacts with the environment to learn and make decisions; (b) The relationship between inputs and outputs in DNN is established through the connections between layers of the neural network; (c) In DRL, leveraging deep neural networks to learn the mapping between states and actions is accomplished through the process of training the neural network.}
    \label{fig:ML}
\end{figure*}

\cref{fig:ML} explains the basic elements of RL, DL, and DRL.
As described in \cref{fig:ML01}, an agent interacts with the environment by observing its state $s_t$ and taking corresponding action $a_t$.
In \cref{fig:ML02}, it illustrates the relationship between input and output, showcasing how deep neural networks adeptly map high-dimensional input data to their corresponding outputs by learning complex mappings and automatically extracting features through advanced representations and patterns.
\cref{fig:ML03} illustrates the method of applying DL within the realm of RL, merging the decision-making optimization capabilities of RL with the representational learning abilities of DL.

From \cref{fig:ML}, we can understand that DRL is a combination of the decision-making capability of DR and the mapping capability of DL. How to perform training using DL and RL will be described in the specific algorithm. This paper adopts the SAC algorithm, a RL algorithm proposed by \citeauthor{pmlrv80haarnoja18b}\cite{pmlrv80haarnoja18b} in 2018. The SAC algorithm is an advanced framework that merges the principles of DRL with maximum entropy theory. It leverages deep neural networks to serve as approximators for value and policy functions, aiming to maximize the entropy of $a_t$ to promote exploration and the development of robust strategies. The SAC algorithm exhibits exceptional performance and applicability in handling problems within continuous action spaces, fostering exploration and the learning of diverse strategies, adaptively adjusting entropy coefficients, and facilitating offline learning\cite{pmlrv80haarnoja18b,DBLP}. Consequently, this paper selects the SAC algorithm to control fluid flow.

The mathematical foundation of RL is encapsulated by the Markov Decision Process (MDP), a model that rigorously defines the dynamics of decision-making environments where outcomes are partly random and partly under the control of a decision maker\cite{puterman1990markov,van2012reinforcement,mao2022active}. An MDP is characterized by a tuple $(S, A, R, P, \rho_0)$, where $S$ represents the set of all possible $s_t$ within the environment. A $s_t$ encapsulates the pertinent information describing the current circumstances of both the environment and the agent. $A$ represents the set of possible $a_t$ that an agent can take. $R: S \times A \rightarrow \mathbb{R}$ is the reward function, providing the immediate reward $r_t$ received after an agent transitions from $s_t$ to $s_{t+1}$ due to $a_t$ at time $t$. $r_t$ is the feedback signal the agent receives based on the $a_t$ it takes and its current $s_t$. $P: S \times A \times S \rightarrow \mathbb{R}$ defines the $s_t$ transition probability function, indicating the likelihood of moving from $s_t$ to $s_{t+1}$ upon taking $a_t$. $\rho_0$ is the initial $s_t$ distribution function, ensuring that the sum of probabilities across all $s_t$ in $S$ equals one ($\sum_{s \in S} \rho_0(s) = 1$).

The MDP framework provides a precise structure for RL problems, defining key elements such as $S$ spaces, $A$ spaces, reward functions, and state transition probabilities. Within this framework, the SAC algorithm introduces entropy regularization, which encourages exploration by maximizing policy entropy. This regularization enhances action diversification, preventing premature convergence to local optima and greatly improving the algorithm's exploration capability. The core optimization goal of the SAC algorithm is as follows\cite{ziebart2010modeling,haarnoja2019soft,pmlrv80haarnoja18b}:

\begin{equation}
\pi^* = \arg\max_{\pi} \sum_t \mathbb{E}_{(s_t,a_t) \sim \rho_\pi} \left[ r(s_t, a_t) + \alpha \mathcal{H}(\pi(\cdot | s_t)) \right],
\end{equation}

where $\pi^*$ denotes the optimal policy, $r(s_t, a_t)$ represents the reward for taking $a_t$ in $s_t$, $\alpha$ signifies the weight of the entropy term in the objective, and $\mathcal{H}(\pi(\cdot | s_t))$ is the entropy of policy $\pi$ given $s_t$. This inclusion of entropy in its objective ensures SAC maintains a balance between exploiting known rewards and exploring novel $a_t$, rendering it highly versatile for a broad spectrum of tasks.

In the SAC algorithm, the state value function \(V(s_t)\) and the action value function \(Q(s_t, a_t)\) are pivotal in determining the optimal policy. These functions facilitate the algorithm's decision-making process by estimating the expected returns from $s_t$ and $a_t$, leveraging the Bellman equation for recursive computation. The SAC algorithm approximates these value functions using DL techniques, enabling efficient handling of complex, high-dimensional environments. Below are the formulations of these value functions within the SAC framework:

\begin{itemize}
    \item \(Q(s_t, a_t)\) estimates the expected return of taking an $a_t$ in $s_t$ and thereafter following the current policy. In the SAC algorithm, \(Q(s_t, a_t)\) is updated based on the Bellman optimality principle, incorporating both the immediate $r_t$ and the discounted future $s_t$ value, adjusted for the entropy of subsequent policy actions. The typical formulation for updating \(Q(s_t, a_t)\) in SAC is:
    \begin{equation}
    Q(s_t, a_t) = r(s_t, a_t) + \gamma \mathbb{E}_{s_{t+1}} [V(s_{t+1})]
    \end{equation}
    where \(r(s_t, a_t)\) is the reward received after executing $a_t$ in $s_t$, \(\gamma\) is the discount factor, and \(\mathcal{E}\) denotes the environment dynamics that govern the transition to the next state \(s_{t+1}\).
    These formulations underscore the SAC algorithm's dual emphasis on maximizing expected returns and promoting policy entropy, thus ensuring a balanced approach to exploration and exploitation in continuous action spaces. The use of DL models to approximate \(V(s_t)\) and \(Q(s_t, a_t)\) allows SAC to effectively navigate and learn from complex, high-dimensional $S$ and $A$ spaces.
    \item \(V(s_t)\) in SAC is aimed at evaluating the expected return of being in $s_t$ and acting according to the current policy \(\pi\). It is defined as the expected value of the \(Q(s_t, a_t)\), minus the product of the temperature parameter \(\alpha\) and the policy's entropy, which encourages exploration by valuing states based not only on immediate $r_t$ but also on the diversity of $a_t$ taken. The formal definition is:
    \begin{equation}
    V(s_t) = \mathbb{E}_{a_t \sim \pi} [Q(s_t, a_t) - \alpha \log \pi(a_t | s_t)]
    \end{equation}
\end{itemize}

In the SAC algorithm, the learning of the \(V(s_t)\), and \(Q(s_t, a_t)\) is facilitated through deep neural networks.\cite{pmlrv80haarnoja18b} These networks utilize experience samples collected from the environment, updating network parameters via the back-propagation algorithm to minimize the loss functions mentioned above. The capability of DL allows SAC to handle high-dimensional $S$ and $A$ spaces, and by employing an end-to-end learning approach, it directly extracts useful features from raw observation data, greatly enhancing the algorithm's generalization ability and applicability.\cite{pmlrv80haarnoja18b,DBLP,haarnoja2019soft}

\subsection{Formulation of the AFC problem using DRL}\label{sec:Formulation of the AFC problem using DRL}

\begin{figure*}[ht]
    \includegraphics[width=0.8\textwidth]{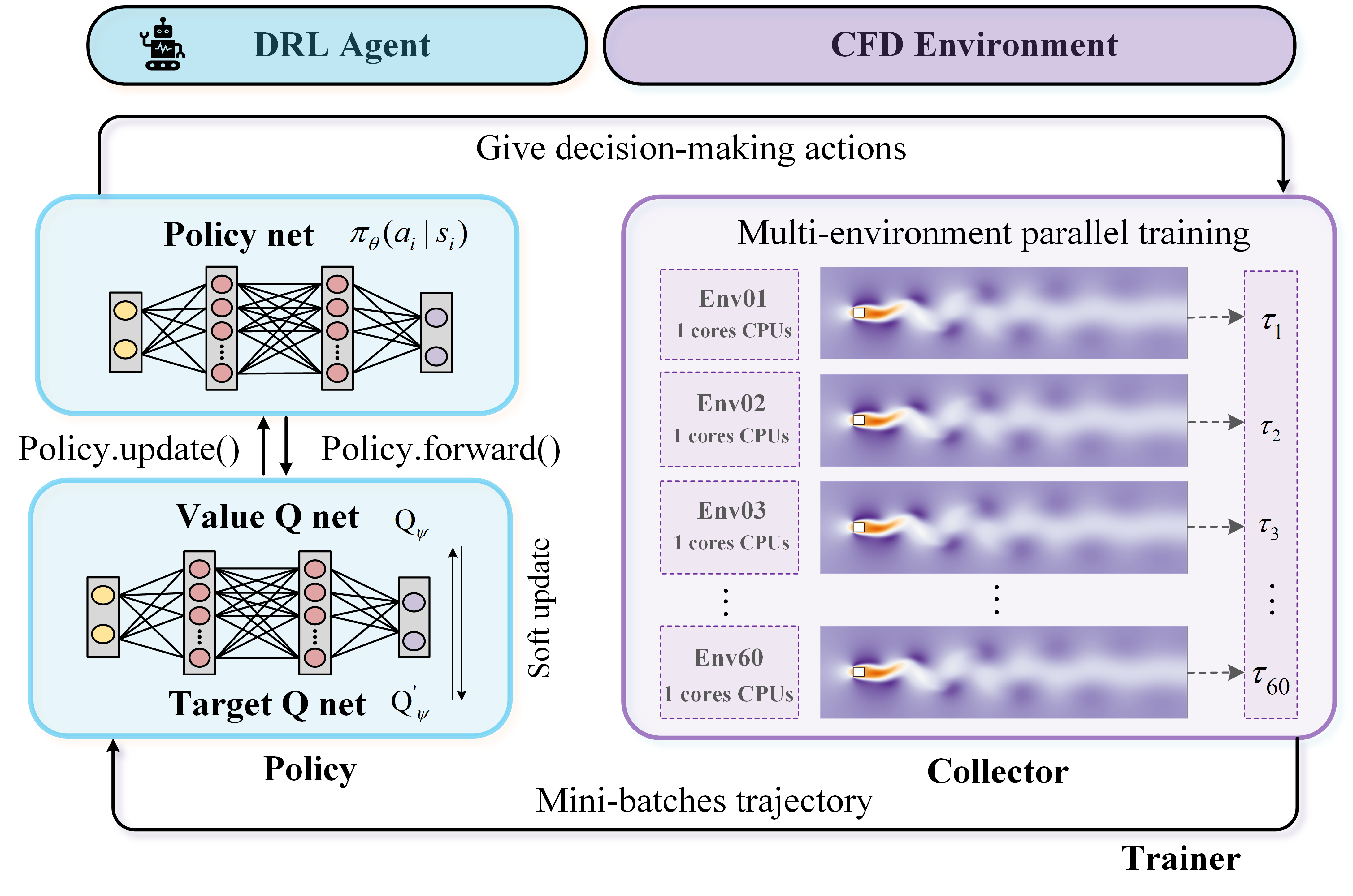}
    \caption{Describes an agent observing states in the environment, taking actions, and obtaining rewards from the environment to learn and optimize its behavior. Posits a scenario with 60 parallel CFD environments, each harnessing the computational power of 1 CPU, cumulatively utilizing 60 cores. These environments are independent CFD environments, capable of concurrently generating 60 separate trajectories to facilitate gradient updates. Neural networks are used to approximate value and policy functions, which are an important part of this computational process.}
    \label{fig:parallerDRL}
\end{figure*}

The study utilizes DRL algorithms to control two synthetic jets placed on two sides of a square cylinder. By adjusting the airflow around the cylinder through suction or blowing from the jets, the research aims to alter the flow field, affecting the velocity distribution and pressure gradient, in order to minimize drag and suppress vortex shedding.
This section outlines how CFD as an environment for DRL algorithms can be coupled with DRL algorithms to solve AFC problems. The integration of CFD into DRL algorithms necessitates an initial comprehension of the fundamental components of DRL algorithms, namely the environment and the agent. Embedding CFD within the DRL algorithm's environment, and utilizing the decision-making capabilities of the agent, aims to achieve the objectives of reducing vortex shedding and drag around a bluff body. This is facilitated through defining key components such as the agent's $s_t$ and $a_t$, and designing a reward function $r_{T_i}$ to complete the agent's configuration. Ultimately, the agent, by observing the current $s_t$ and selecting corresponding $a_t$, obtains $r_t$ based on the feedback from the environment. The environment, in turn, transitions to a subsequent $s_t$ based on the agent's $a_t$ and the current $s_t$, providing corresponding $r_t$. This iterative process underpins the synergy between CFD and DRL, offering a novel approach to optimizing fluid dynamics problems through learned behaviors. The design of the fundamental framework, as delineated in \cref{fig:parallerDRL}, draws upon the collective works of \citeauthor{rabaultArtificial}\cite{rabaultArtificial},\citeauthor{rabault2020deep}\cite{rabault2020deep}, \citeauthor{wangDRLinFluids}\cite{wangDRLinFluids}, \citeauthor{liReinforcementlearning}\cite{liReinforcementlearning}, \citeauthor{parisRobustFlowControl2021a}\cite{parisRobustFlowControl2021a} and \citeauthor{jia2024optimal}\cite{jia2024optimal}.

\begin{itemize}
    \item $s_t$: At each temporal juncture, the $s_t$ is constituted by instantaneous flow field or pressure data collected from specific locations within the computational domain of the numerical simulation. Around the bluff body and in the wake region, 201 probes are strategically placed to capture the transient states of the CFD environment. The placement of these probes is informed by the research findings of \citeauthor{wangDRLinFluids}\cite{wangDRLinFluids}, \citeauthor{liReinforcementlearning}\cite{liReinforcementlearning}, \citeauthor{parisRobustFlowControl2021a}\cite{parisRobustFlowControl2021a}, ensuring comprehensive coverage around the bluff body and within the wake region. This arrangement scrutinizes the placement of probes to ensure they encompass the maximum values and boundary locations of velocity fields, pressure fields, and their fluctuating counterparts, thereby capturing key flow characteristics essential for accurate analysis and decision-making within the DRL framework. 
    
    \item $a_t$ : the $a_t$ of the agent are defined as the velocity of the synthetic jets, with the stipulation that the magnitude of these action cannot exceed 2\% of the inlet velocity. This constraint is designed with the intention of achieving drag reduction using the smallest possible jet velocity, thus ensuring the energy efficiency of the AFC technology. To guarantee the smoothness of continuous $a_t$ provided by the agent to the environment, a smoothing function is employed between adjacent $a_t$. Let $a_t$ and \(a_{t+1}\) represent the $a_t$ magnitudes at consecutive time steps. The smoothing function \(S\) applied between these $a_t$ can be defined by a mathematical relationship that ensures a gradual transition from $a_t$ to \(a_{t+1}\), thereby mitigating abrupt changes in the jet velocity. Smoothing function \(S\) can be defined as:
    \begin{equation}
    S(V_{\Gamma_1,T_i}, a, V_{\Gamma_1,T_{i-1}}) = V_{\Gamma_1,T_i} + \beta\cdot(a - V_{\Gamma_1,T_{i-1}}),
    \end{equation}
    where \(V_{\Gamma_1,T_i}'\) is the updated value at time step \(i\), \(V_{\Gamma_1,T_i}\) is the current value at time step \(i\), \(a\) represents the target $a_t$ magnitude, \(V_{\Gamma_1,T_{i-1}}\) is the value at the previous time step \(i-1\), and \(\beta\) is a coefficient determining the extent of adjustment towards the target \(a\). This function effectively interpolates between the previous value and the target $a_t$, with \(\beta\) controlling the smoothness of the transition.

    \item $r_t$ : The design of the reward function is crucial for guiding effective learning, particularly in fluid dynamics applications like reducing drag around a bluff body, where selecting appropriate performance metrics is key. In this study, the reward function incorporates both the $C_D$ and $C_L$, which are critical indicators of performance in fluid flow past objects. 
    
    The specific reward function takes the following form:
    \begin{equation}\label{eq:my reward}
        r_{T_i}=C_{D,0}-\left(C_D\right)_{T_i}-\omega\left|\left(C_L\right)_{T_i}\right|.
    \end{equation}
    \( C_{D,0} \) represents the baseline $C_D$, serving as a reference point for drag coefficient. The addition of the reference point is to drive the overall reward to positive values, which helps with the convergence during training. \( (C_D)_{T_i} \) denotes the $C_D$ at time step \(T_i\), with the objective of minimizing this value relative to the baseline. \( (C_L)_{T_i} \) signifies the $C_L$ at time step \(T_i\), whose absolute value is penalized to mitigate lift forces that may destabilize the flow around the bluff body. \( \omega \) is a weighting factor that quantifies the trade-off between minimizing drag and controlling lift fluctuations. Since the flow we are interested in is highly directional, the weighting factor is typically between 0.1 and 0.2, indicating that drag reduction and lift reduction are valued differently. In this study, the exact value of hyperparameter \( \omega \) is tuned for performance by following the advice given by \citeauthor{rabaultArtificial}\cite{rabaultArtificial}.

\end{itemize}

In this study, numerical simulations were conducted with a time step size of \(5 \times 10^{-4}\) seconds, and the duration of the control time step was set to 0.025 seconds, equivalent to 25 numerical simulation time steps. Consequently, the total duration of a training dataset comprising 100 steps amounts to 1.25 seconds, which corresponds to 2500 numerical simulation time steps. In scenarios with $Re$ ranging from 100 to 400, the vortex shedding period is approximately between 0.174 and 0.191 seconds. Thus, one epoch is designed to last for 1.25 seconds, equating to about 6.6 to 7.2 vortex shedding periods. Each training epoch encompasses several vortex shedding periods, allowing the agent to thoroughly observe and adapt to the flow dynamics within multiple shedding periods.

\subsection{Parallelization of DRL-based AFC problem}\label{sec:Parallelization of DRL-based AFC problem} 

This parallel strategy draws inspiration from the research conducted by \citeauthor{rabaultAccelerating}\cite{rabaultAccelerating} and \citeauthor{wangDRLinFluids}\cite{wangDRLinFluids}, while also incorporating insights from the work of \citeauthor{jia2024optimal}\cite{jia2024optimal}. For a more comprehensive understanding, we encourage readers to consult their work and the references therein. 
The DRL training and CFD simulations presented in this study were performed on a high-performance computing system. The computational framework employed for the analyses is powered by an \texttt{Intel\textsuperscript{\textregistered} Xeon\textsuperscript{\textregistered} Platinum 8358 CPU}, operating at 2.60GHz. This system is fortified with a total of 64 cores, distributed evenly across two sockets, with each socket housing 32 cores. The CFD were conducted utilizing \texttt{OpenFOAM\textsuperscript{\textregistered}} version 8
\cite{jasakOpenFOAMLibraryComplex2013}, an acclaimed open-source software platform. 
For the DRL aspect, the open-source Python framework \texttt{Tianshou}\cite{weng2021tianshou} was employed.
\cref{fig:parallerDRL} illustrates the allocation and utilization of CPU cores during the coupled training of DRL algorithms and CFD simulations, showcasing the fundamental parallelization strategy.

\section{Results and discussion}\label{sec:Results and discussion}

\cref{sec:sec1} details the baseline fluid dynamics around a square cylinder for various $Re$ (100, 200, 300 and 400), with a focus on the examination of fluid dynamic parameters such as $C_D$, $C_L$, and $St$. The analysis furnishes a reference for selecting pertinent coefficients for AFC strategies and lays the groundwork for a comparative assessment of control effectiveness.
In \cref{sec:sec2}, the SAC algorithm is employed to evaluate the control performance and reliability of applying AFC techniques on the flow around a square cylinder at various $Re$. The primary focus is to ascertain whether the DRL agent can effectively manipulate vortex shedding in flow configurations at increasing $Re$ to achieve a reduction in drag and control over lift. 
\cref{sec:sec3} discusses detailed comparison of flow fields before and after control application, revealing the mechanisms behind the algorithm's effectiveness in modifying fluid behavior and reducing flow separation.

\subsection{Flow Analysis of a Square Cylinder at Different Reynolds Numbers}\label{sec:sec1}

\begin{figure*}
\includegraphics[width=0.9\textwidth]{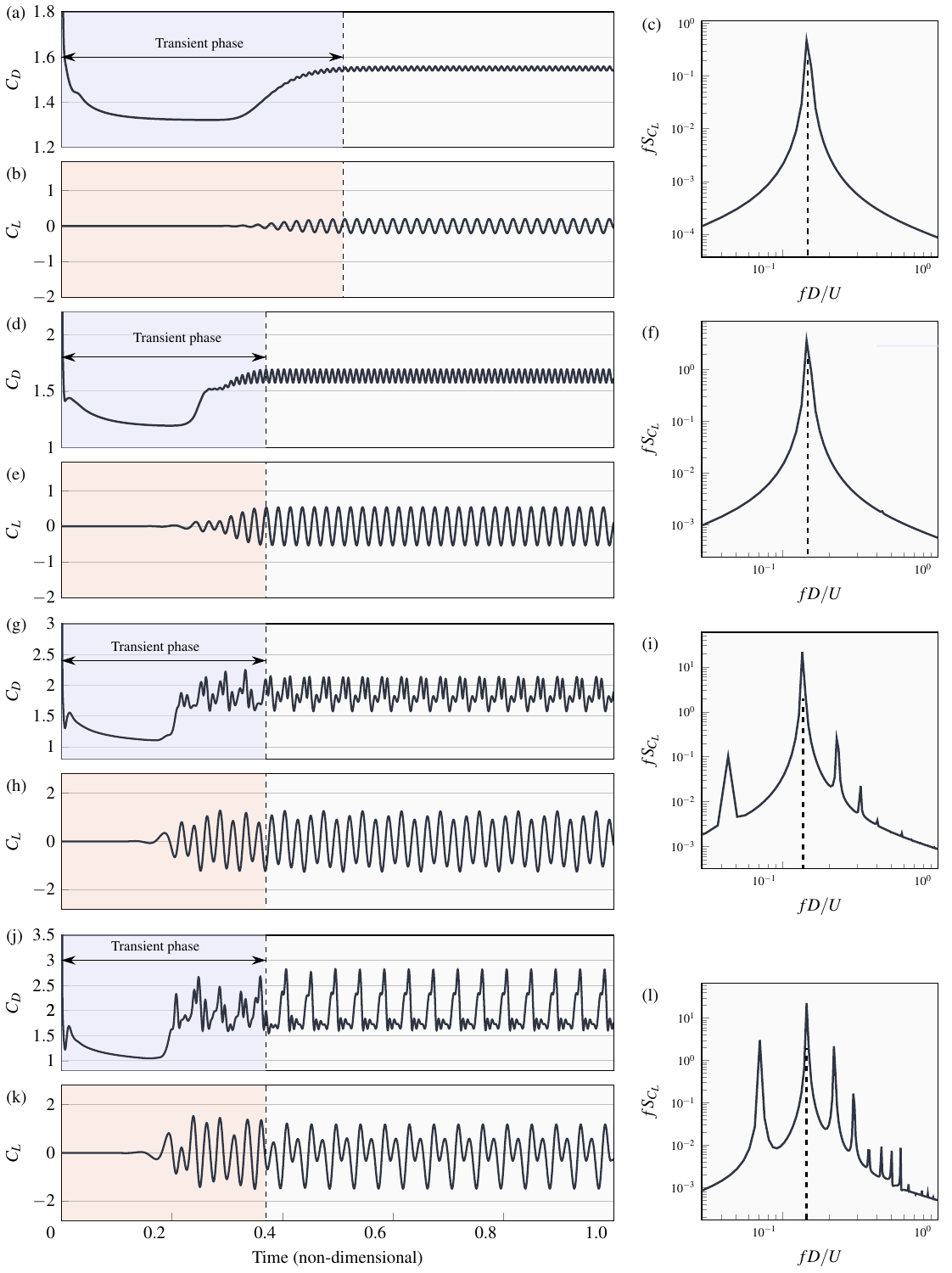}
\caption{Unsteady flow past a square cylinder. For $Re = 100$ (a) time series of $C_D$; (b) time series of $C_L$; (c) PSD of $C_L$ ; For $Re = 200$ (d) time series of $C_D$; (e) time series of $C_L$; (f) PSD of $C_L$. For $Re = 300$ (g) time series of $C_D$; (h) time series of $C_L$; (i) PSD of $C_L$; For $Re = 400$ (j) time series of $C_D$; (k) time series of $C_L$; (l) PSD of $C_L$.}
\label{fig:baseline1}
\end{figure*}

The analysis of the flow around a square cylinder at $Re = 100$, $Re = 200$, $Re = 300$ and $Re = 400$ is depicted in \cref{fig:baseline1}, which present the variation of $C_D$, $C_L$, and 
PSD (Power Spectral Density) of $C_L$ over time. 
Starting with $Re = 100$, Fig.5 (a) elucidates the temporal evolution of the $C_D$. Initially, there is a marked transient phase where $C_D$ exhibits a gradual decrease, then stabilizing to a quasi-steady state as the flow regime reaches equilibrium. This behavior is indicative of the flow adjustment from an initial transient response to a stable vortex shedding pattern. In Fig.5 (b), following a transition phase, the $C_L$ exhibits periodic oscillations within a magnitude of $\pm 0.2$. These oscillations are symmetric about the zero baseline, indicating that the lift acting on the cylinder has an alternating character due to the periodic shedding of vortices.
Fig.5 (c) is the PSD of the $C_L$. The non-dimensionalized frequency $fD/U = 0.142$ corresponding to the peak value reflects the main vortex shedding frequency, indicating the periodicity of the lift exerted on the object due to the alternating separation of the vortex.
At a $Re$ of 200, the development patterns of drag and lift around a square cylinder are fundamentally similar to those observed at $Re = 100$. However, as depicted in Fig.5 (d), the transition to a stable phase of periodic vortex shedding around the square cylinder occurs more rapidly at $Re = 200$. Correspondingly, as shown in Fig.5 (e), the $C_L$ of the square cylinder also progresses through a transition phase before entering a periodic stage, where the magnitude of $C_L$ oscillations is greater compared to the scenario at $Re = 100$. Similarly, Fig.5 (f) illustrates a PSD with a distinct peak, indicating that the phenomenon of vortex shedding is dominated by a single frequency, and it is consistent to the shedding frequency for $Re = 100$.

As we increase the $Re$ to 300, the $C_D$ around the square cylinder, as illustrated in Fig.5 (g), transitions at a faster rate from an initial transient stage to a phase characterized by periodic oscillations. The periodicity of the $C_D$ oscillations at $Re = 300$ is significantly different from that observed at $Re = 100$ and $Re = 200$.
In the stable phase of periodic oscillations at $Re = 300$, the drag curve exhibits multiple distinct extremal points within each period. Correspondingly, in Fig.5 (h), the $C_L$ also displays multiple oscillation amplitudes, oscillating not at a singular frequency but rather at multiple frequencies that are harmonics of the lowest peak frequency. The PSD calculations for the lift, as shown in Fig.5 (i), further corroborate that the periodic oscillations of lift are governed by multiple frequencies. The largest spectral density still peaks close to the frequency at $Re = 100$ or $200$. However, the emergence of harmonics indicate the onset of multi-scale in the underlying flow.
When the $Re$ reaches 400, both the amplitude of the $C_D$ (Fig.5 (j)) and $C_L$ (Fig.5 (k)) surpass those observed at other $Re$, accompanied by a more complex fluctuation frequency spectrum. 
The power spectral density graph reveals even more harmonics at higher frequencies, shown in Fig.5 (l),indicating the predominance of small-scale vortices at higher frequencies in the vortex shedding phenomenon. 

Overall, through the analysis of the characteristics of $C_D$, $C_L$ and $f$ across four different $Re$, it is observed that at $Re = 100$ and $Re = 200$, the flow around the square cylinder is characterized by a vortex shedding pattern dominated by a single frequency. Conversely, at $Re = 300$ and $Re = 400$, the vortex shedding pattern around the square cylinder becomes more chaotic and multiple harmonic frequencies emerge. This transition indicates an increase in flow complexity as the $Re$ rises, implying that flow control encounters greater challenges at higher $Re$.

\subsection{Control effect based on SAC algorithm}\label{sec:sec2}

\begin{figure*}[hbtp]
\centering
    \begin{subfigure}{0.495\textwidth}
    \includegraphics[width=\textwidth]{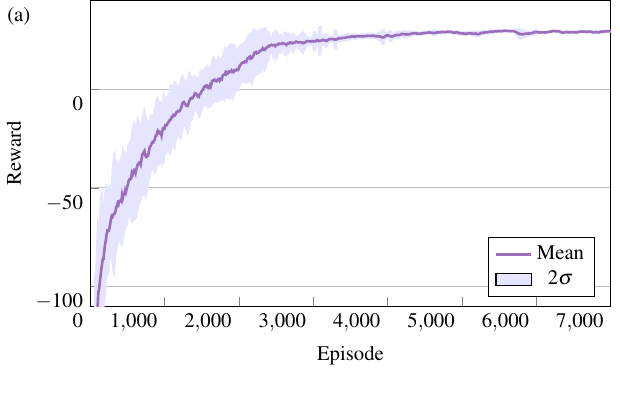} 
    \caption{}
    \label{fig:reward_re100}
    \end{subfigure}  
    \begin{subfigure}{0.495\textwidth}
    \includegraphics[width=\textwidth]{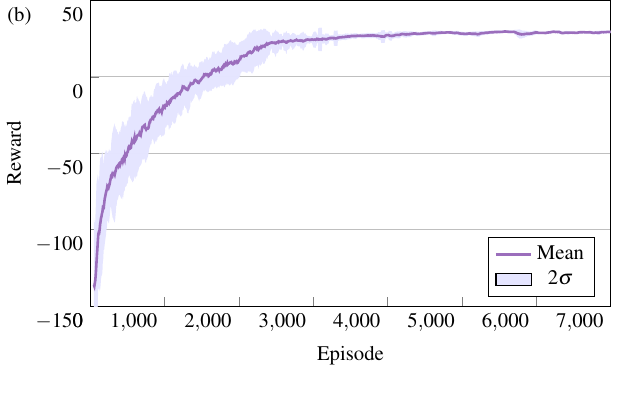} 
    \caption{}
    \label{fig:reward_re200}
    \end{subfigure}  
    \begin{subfigure}{0.495\textwidth}
    \includegraphics[width=\textwidth]{0601figure.pdf} 
    \caption{}
    \label{fig:reward_re300}
    \end{subfigure}      
    \begin{subfigure}{0.495\textwidth}
    \includegraphics[width=\textwidth]{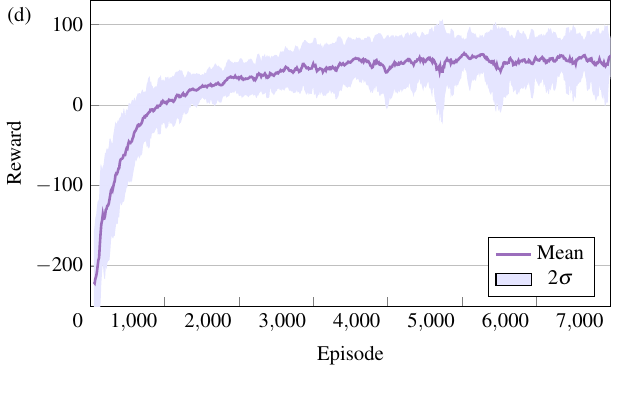} 
    \label{fig:reward_re400}
    \end{subfigure}  
    \caption{Cumulative reward for the DRL training process using the SAC algorithm. The solid curves depict the mean reward at each episode, with the shaded area representing a range of two standard deviations ($2\sigma$) from the mean. (a) $Re=100$; (b) $Re=200$; (c) $Re=300$; (d) $Re=400$.}
    \label{fig:reward}
\end{figure*}

In the previous section, we observed that at $Re = 300$ and $Re = 400$, wake vortex shedding behind the square cylinder exhibits a complex pattern dominated by multiple frequencies, in stark contrast to the periodic, single-frequency dominated shedding observed at $Re = 100$ and $Re = 200$.
Although \citeauthor{wangDRLinFluids}\cite{wangDRLinFluids} has demonstrated that an DRL algorithm can achieved a drag reduction effect of 13.7\% at $Re = 100$, the vortex shedding phenomenon still prevails. Whether an effective control of the wake vortex can be achieved at high $Re$ still remain elusive.
Therefore, this section aims to verify whether active flow control based on an DRL algorithm can intelligently identify control strategies at a range of $Re$ to suppress the shedding of wake vortices, thus reducing drag and mitigating lift oscillations.


We use the SAC algorithm to implement active flow control, and use a pair of synthetic jets to control the flow field of a square cylinder. \cref{fig:reward} presents the reward function profiles from the DRL training process using the SAC algorithm across $Re = 100$, 200, 300 and 400 respectively. As depicted in \cref{fig:reward_re100}, at $Re = 100$, the SAC agent's learning trajectory is initially characterized by lower rewards, yet it exhibits rapid improvement within the initial episodes. The learning curve swiftly stabilizes, culminating in a plateau around the 2,500th episode, indicative of a consolidated learning outcome. The compact $2\sigma$ interval observed in later stages signifies consistent agent performance, implying a stable and convergent training process at this $Re$. Based on the definition of the reward function in \cref{eq:my reward}, a positive reward value indicates a successful reduction in the mean drag, suggesting that the agent have learned from simulation data to achieve the objective at $Re = 100$.

At higher $Re$, similar learning trajectories are evident in \cref{fig:reward_re200,fig:reward_re300,fig:reward_re400}, where rewards in each instance converge over time. This consistency underscores the remarkable proficiency of DRL agents in acquiring effective control methods across a spectrum of $Re$. Beyond this commonality, it is observable that the extent of variation within each reward trajectory differs. With an increase in the $Re$, the level of fluctuation intensifies. Specifically, at $Re = 400$, the variability remains comparatively large at a notable magnitude even while the average reward stabilizes. This extensive fluctuation in the reward signal at higher $Re$ suggests that more complex flow dynamics present greater challenges, necessitating more sophisticated control strategies. 
 
\begin{figure*}[ht]
\centering
\includegraphics[width=\textwidth]{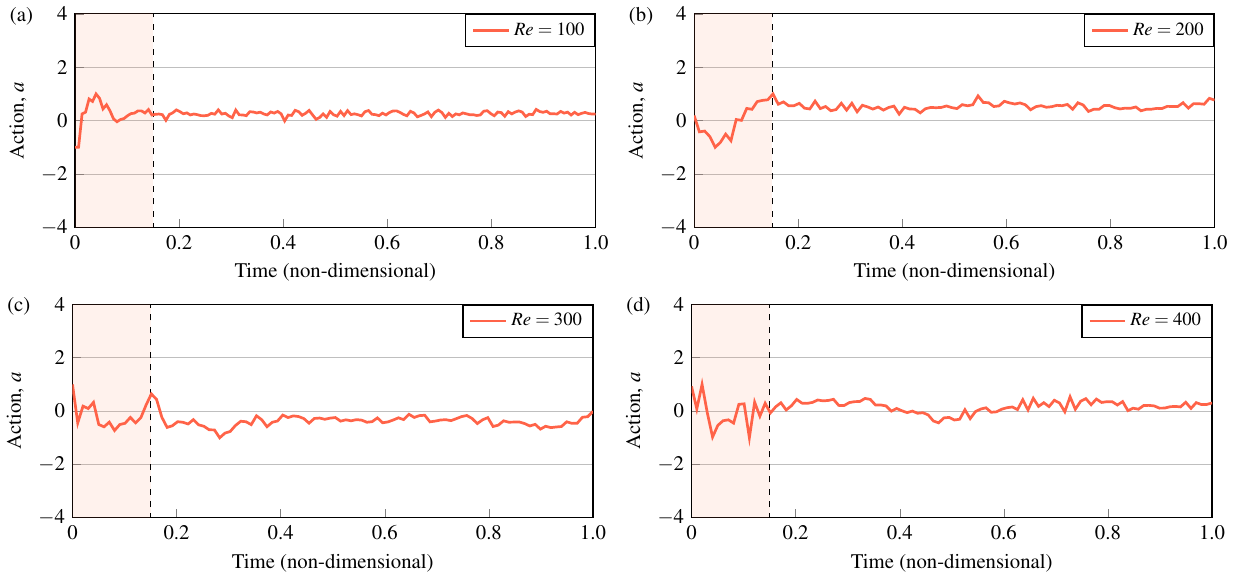}
\caption{ The value of the action generated during the interaction between the agent and the environment is the value of the mass flow rate of the synthetic jet. (a) $Re=100$; (b) $Re=200$; (c) $Re=300$; (d) $Re=400$.}
\label{fig:Reaction}
\end{figure*}


\begin{figure*}[htbp]
\centering
\includegraphics[width=0.95\textwidth]{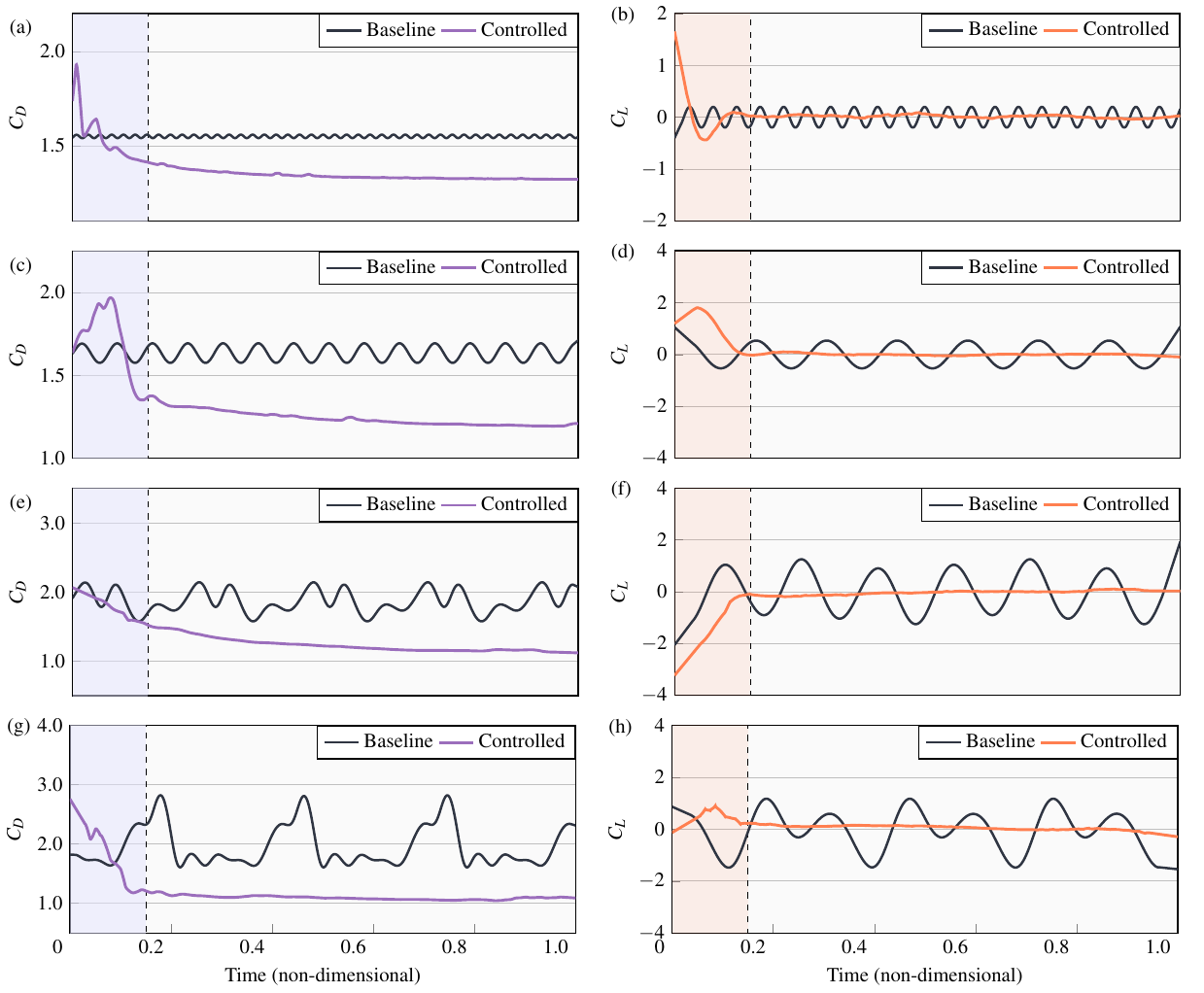}
\caption{Comparative diagram of the fluid force coefficients before and after flow control. The left side is $C_D$ and the right side is $C_L$. (a) and (b) $Re=100$; (c) and (d) $Re=200$; (e) and (f) $Re=300$; (g) and (h) $Re=400$.}
\label{fig:drlcdclre}
\end{figure*}

\cref{fig:Reaction} depicts the continuous adjustment of jet velocity by the SAC agent during its interaction with the CFD environment, across $Re=100$, 200, 300, and 400, respectively. 
In \cref{fig:drlcdclre}, the consequent variations in $C_D$ and $C_L$ are presented. The reference baseline is included included to enable comparisons between the flow conditions before and after the implementation of control. 
The initial phase of control is characterized by a pronounced surge in jet velocity, which subsequently undergoes a gradual decline before achieving stabilization near a negligible value. 
Under the influence of the synthetic jet, both the $C_D$ and $C_L$ fluctuate significantly as well, then sharply decrease and eventually reach a stable equilibrium. Notably, across all conducted experiments, there is a substantial reduction in the $C_D$, while the $C_L$ approaches a value near zero. Even at higher $Re$ where the baseline flow is more chaotic, the DRL algorithm succeeds in devising effective flow control strategies that mitigate drag and lift fluctuations.
The near-zero magnitude of the jet velocity also implies that minimal energy is required to sustain the stabilized flow for long-duration drag reduction.

\begin{table*}[htbp]
\centering
\begin{threeparttable}
\caption{Summary of results for our work and related research on AFC applied to square cylinder using DRL.}
\vspace{-\baselineskip}
\begin{tabularx}{\textwidth}{
  >{\centering\arraybackslash}p{0.06\linewidth}
  >{\centering\arraybackslash}p{0.1\linewidth}
  >{\centering\arraybackslash}p{0.1\linewidth}
  >{\centering\arraybackslash}p{0.1\linewidth}
  >{\centering\arraybackslash}p{0.13\linewidth}
  >{\centering\arraybackslash}p{0.1\linewidth}
  >{\centering\arraybackslash}p{0.1\linewidth}
  >{\centering\arraybackslash}p{0.12\linewidth}
  >{\centering\arraybackslash}p{0.13\linewidth}  
}
\toprule
$Re$ & Jet Location & $\overline{C}_{D,\text{Baseline}}$ & $\overline{C}_{D,\text{Controlled}}$ & Drag reduction rate (\%) & $\overline{C}_{L,\text{Controlled}}$ & $Action$, $\overline{a}$ & Vortex shedding suppression & Reference \\
\hline
\hline
\multirow{4}{*}{100} &  Trailing & 1.549 & 1.337 & 13.7 & - & - & NO & \citeauthor{wangDRLinFluids}\cite{wangDRLinFluids} \\
 & Leading   & 1.476 & 1.370 & 7.2  & -  & -  & NO & \citeauthor{chen2023deep}\cite{chen2023deep} \\
 &  Middle   & 1.476 & 1.440 & 2.4  & - & - & NO & \citeauthor{chen2023deep}\cite{chen2023deep} \\
 &  Trailing & 1.476 & 1.280 & 13.3 & - & -  & YES & \citeauthor{chen2023deep}\cite{chen2023deep} \\   
\midrule
100  & Trailing  & 1.549 & 1.325 & 14.4  & 0.0013  & 0.2675  & YES &  \multirow{4}{*}{Present study}\\
200 & Trailing & 1.633 & 1.203 & 26.4 & -0.0101 & 0.1180 & YES &  \\

300 & Trailing & 1.854 & 1.133 & 38.9 & -0.0240 & -0.3655 & YES & \\

400 & Trailing & 1.964 & 1.041 & 47.0 & -0.0065 & 0.05905 & YES & \\
\bottomrule
\end{tabularx}
\begin{footnotesize} 
\begin{tablenotes}[flushleft]
\item[] a. Jet Location: synthetic jets are located on the lateral or top/bottom sides of the square cylinder. Specifically, 'Trailing' refers to the rear corners; 'middle' refers to the middle position between front and rear sides; and 'leading' refers to the front corners of the square cylinder.
\item[] b. $\overline{C}_{D,\text{Baseline}}$ represents the time-averaged drag of the square cylinder after the flow field stabilizes without AFC. $\overline{C}_{D,\text{Controlled}}$, $\overline{C}_{L,\text{Controlled}}$, and $\overline{a}$ denote the time-averaged values under stable conditions after AFC is applied, excluding the initial transient phase.
\item[] c. "Vortex shedding suppression" indicates whether synthetic jets control can fully prevent vortex shedding in the wake of the square cylinder. Please refer to \cref{fig:velocity_contour} for flow visualizations.
\end{tablenotes}
\end{footnotesize}
\label{tab:01dragreduction}
\end{threeparttable}
\end{table*}

We summarize the results of active flow control performed at multiple Reynolds numbers in \cref{tab:01dragreduction}. The quantities of interest include the reduction ratio of time-averaged drag, time-averaged lift, time-averaged action values, and the suppression of vortex shedding after control. We delay the discussion of vortex shedding suppression to \cref{sec:sec3}. We also include published results from existing literature. 
For a $Re$ of 100, our study achieves approximately 14.4\% reduction in average drag, surpassing the drag reduction reported by \citeauthor{wangDRLinFluids}\cite{wangDRLinFluids} and \citeauthor{chen2023deep}\cite{chen2023deep}. It is worth mentioning that \citeauthor{chen2023deep}\cite{chen2023deep} found trailing jets to be more effective than leading jets, and they were able to achieve vortex shedding suppression with trailing jet control. At $Re$ values of 200, 300, and 400, the AFC strategy implemented in our study achieves drag reduction rates of 26.4\%, 38.9\%, and 47.0\% respectively, while maintaining the lift coefficient near zero. 
Furthermore, the action value of the synthetic jet evidences the strategy's compliance with the low-energy consumption requirements inherent in AFC technologies.

\begin{figure*}[htbp]
\includegraphics[width=0.95\textwidth]{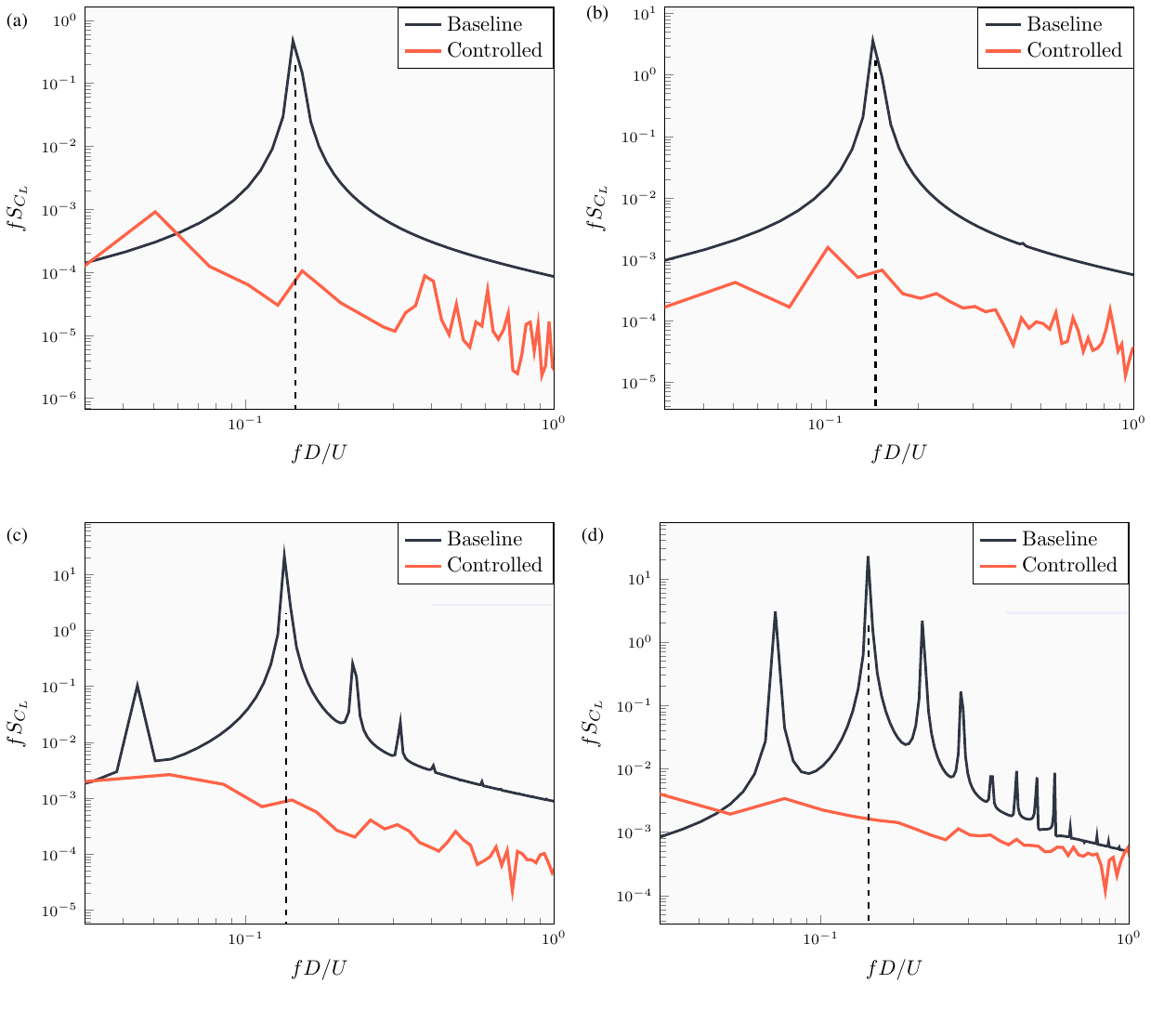}
\caption{A comparative diagram of the PSD of $C_L$ before and after flow control. (a) $Re=100$; (b) $Re=200$; (c) $Re=300$; (d) $Re=400$.}
\label{fig:09drlpsd}
\end{figure*}

In \cref{fig:09drlpsd}, we compare the PSD in the aerodynamic coefficients before and after flow control. In the baseline scenario without control, the PSD shows prominent peaks at specific frequencies near $fD/U = 0.142$, indicating the periodic nature of the fluid vibrations and the corresponding dominant frequencies. Conversely, upon the implementation of flow control, these prominent peaks are either remarkably attenuated or completely vanish, suggesting that the control strategy has significantly altered the oscillatory behavior of the fluid and effectively suppressed periodic fluctuations. At $Re$ of 100 and 200 , the previously significant peaks in the PSD graphs are eliminated after control, implying that the flow control measures have disrupted the periodic fluctuations of the $C_L$. 
For $Re$ of 300 and 400 , multiple significant frequency peaks are evident in the baseline state without control. However, after the application of flow control, all characteristic frequencies are eliminated as well, indicating that the DRL-based control strategy has a mitigating effect on multiple frequencies at higher $Re$. This highlights the success and robustness of the SAC agent in identifying relevant frequencies and deploying intelligent control strategies to mitigate drag and regulate lift fluctuations. The capability to react to multi-scale frequencies provides promising aspects of the adaptability of DRL in chaotic and turbulent AFC scenarios.

\begin{figure*}[htbp]
\centering
\includegraphics[width=\textwidth]{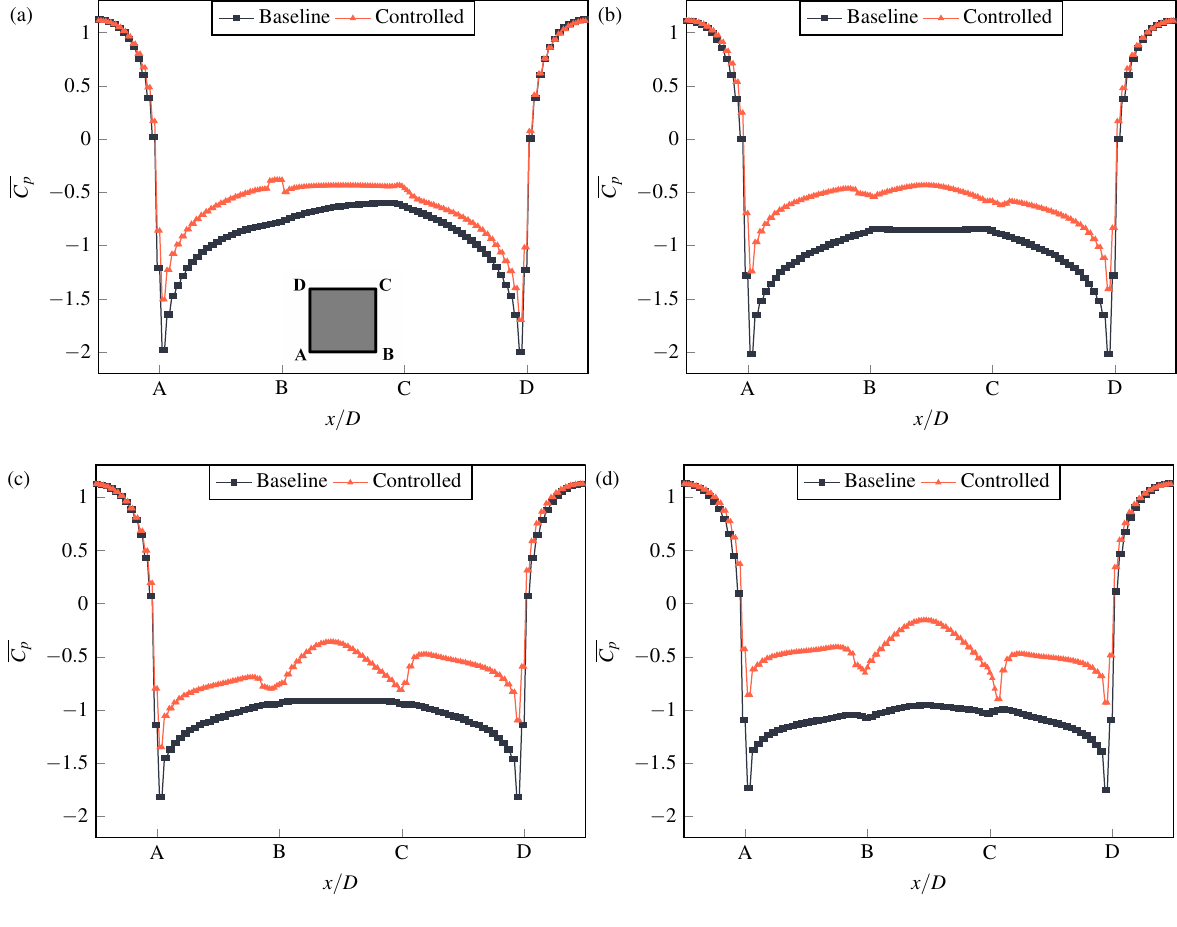}
\caption{Illustration of the training process: profiles of the average pressure coefficient. (a) $Re=100$; (b) $Re=200$; (c) $Re=300$; (d) $Re=400$.}
\label{fig:pressure}
\end{figure*}

Lastly, \cref{fig:pressure} illustrates the comparison between the time-averaged pressure distributions on the surface of the square cylinder under both controlled and uncontrolled flow scenarios. The square cylinder possesses a side length denoted by \(D\), with its center located at the origin of the coordinate system. The specific points on the square cylinder are designated as follows: point A at \((-D/2, -D/2)\), point B at \((D/2,-D/2)\), point C at \((D/2, D/2)\), and point D at \((-D/2, D/2)\). 
In Fig.10 (a), At $Re = 100$, the control strategy deployed by the SAC agent has a negligible effect on the pressure distribution along the AD edge (the windward face) of the square cylinder. Conversely, it substantially affects the pressure distribution on both lateral sides (AB and CD) and the leeward face (BC) of the cylinder, leading to a marked decrease in the average pressure at these sites following the implementation of control measures.
Similarly, Fig.10 (b) illustrates that at $Re = 200$, the changes in the average pressure distribution around the square cylinder before and after control are essentially consistent with those observed at $Re = 100$. Notably, at $Re = 200$, the reduction in pressure distribution on both lateral sides and the leeward face of the cylinder is even more pronounced.
In Fig.10 (c) and Fig.10 (d), under conditions of $Re = 300$ and $Re = 400$ the flow control strategy continues to exert minimal influence on the average pressure distribution on the windward face of the square cylinder. Similar to observations at \(Re = 100\) and \(Re = 200\), the impact on the pressure distribution on both lateral sides and the leeward face of the cylinder is more significant. Notably, the influence of flow control on the pressure distribution on the leeward face is greater than that on the lateral sides, with the most pronounced effect occurring near the middle of the leeward face. 
Overall, this set of comparisons illustrate that the majority of drag reduction is achieved via pressure forces, which highlights the significance of suppressing vortical structures for optimal performance.

\subsection{Detailed flow field comparison before and after control}\label{sec:sec3}

\begin{figure*}[hbtp]
    \centering
    \begin{subfigure}{0.3\textwidth}
    \includegraphics[width=\textwidth]{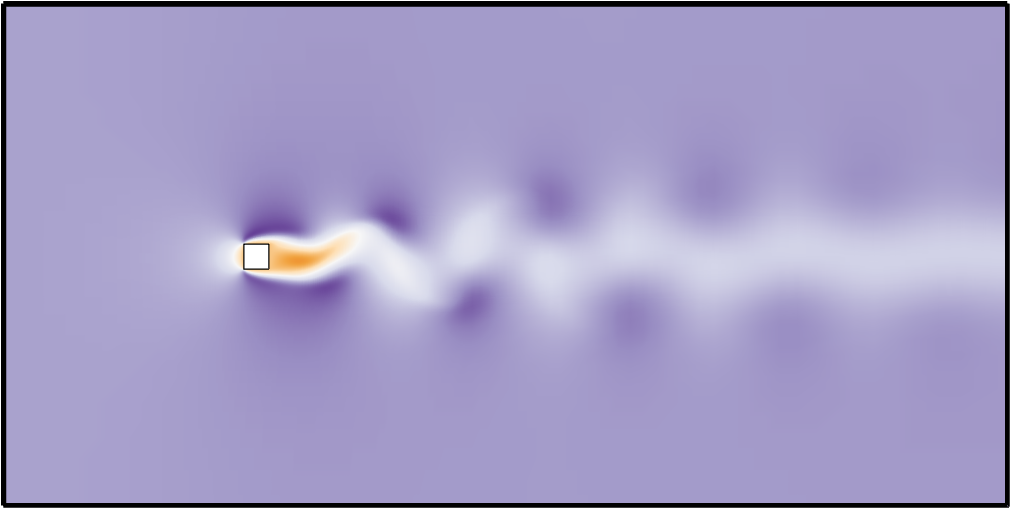}
    \caption{$Re=100$, $t_0$}
     \label{fig:1101Re100}
    \end{subfigure}
    \begin{subfigure}{0.3\textwidth}
    \includegraphics[width=\textwidth]{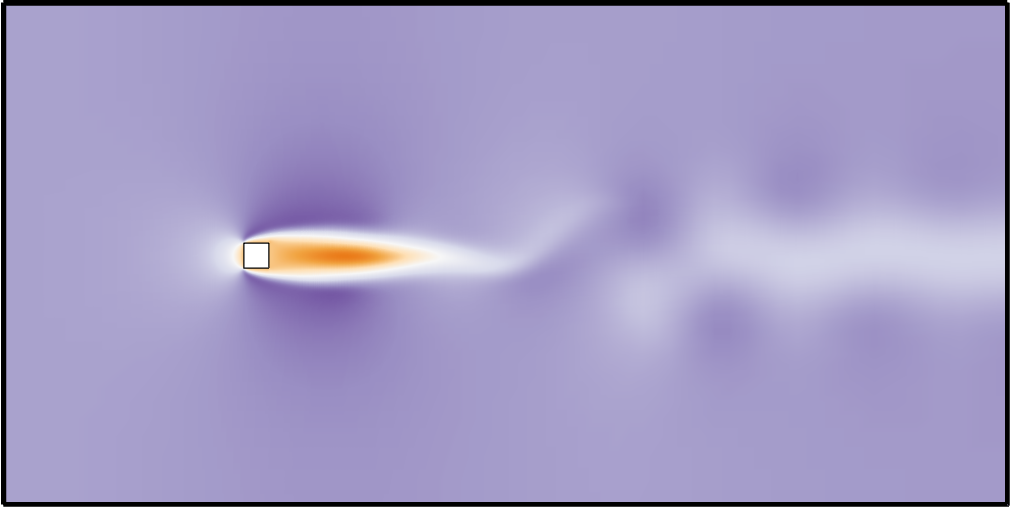}
    \caption{$Re=100$, $t_1$}
    \label{fig:1102Re100}
    \end{subfigure}
    \begin{subfigure}{0.35\textwidth}
    \includegraphics[width=\textwidth]{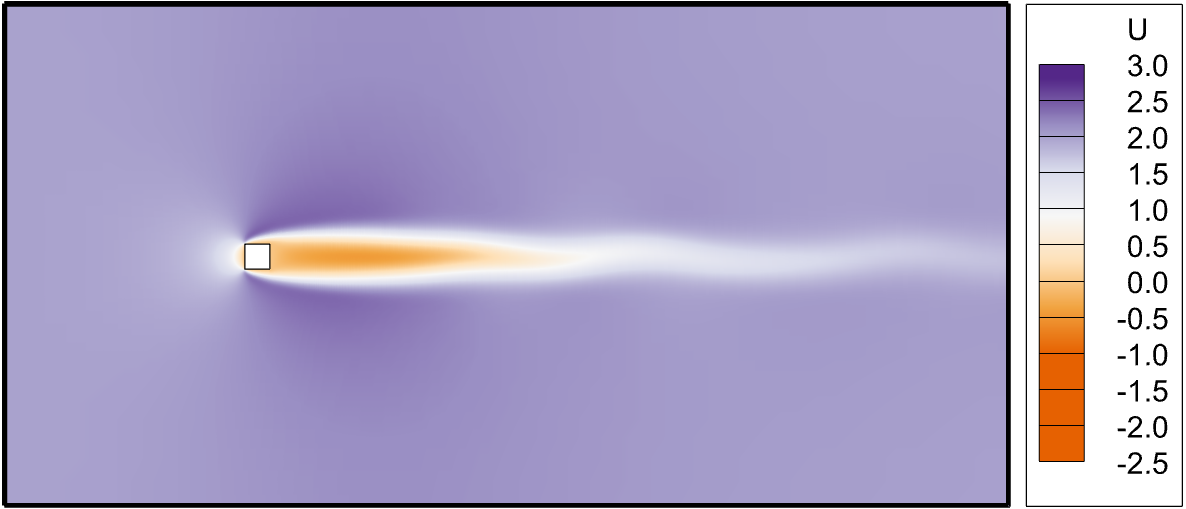}
    \caption{$Re=100$, $t_2$}
    \label{fig:1103Re100}
    \end{subfigure}
    
    \begin{subfigure}{0.3\textwidth}
    \includegraphics[width=\textwidth]{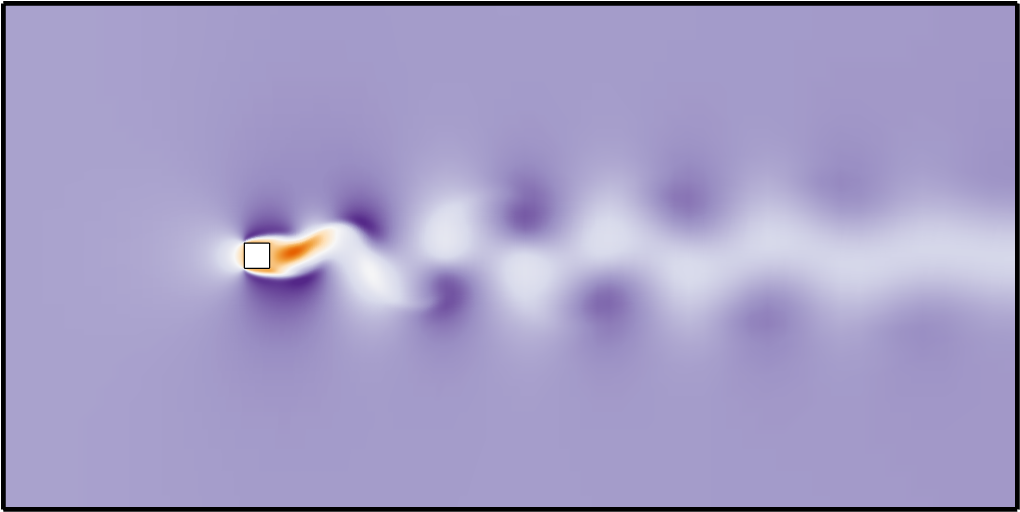}
    \caption{$Re=200$, $t_0$}
    \label{fig:1104Re200}
    \end{subfigure}
    \begin{subfigure}{0.3\textwidth}
    \includegraphics[width=\textwidth]{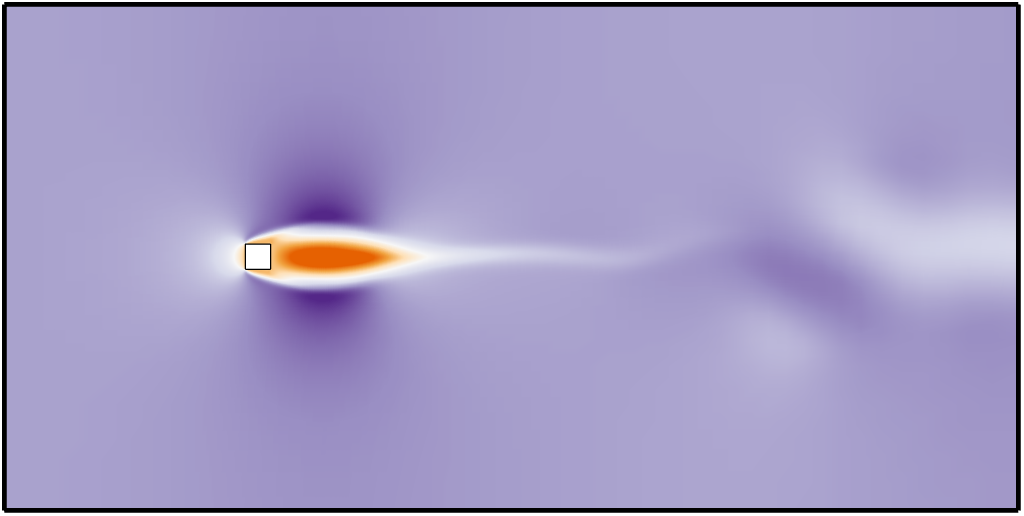}
    \caption{$Re=200$, $t_1$}
    \label{fig:1105Re200}
    \end{subfigure}
    \begin{subfigure}{0.35\textwidth}
    \includegraphics[width=\textwidth]{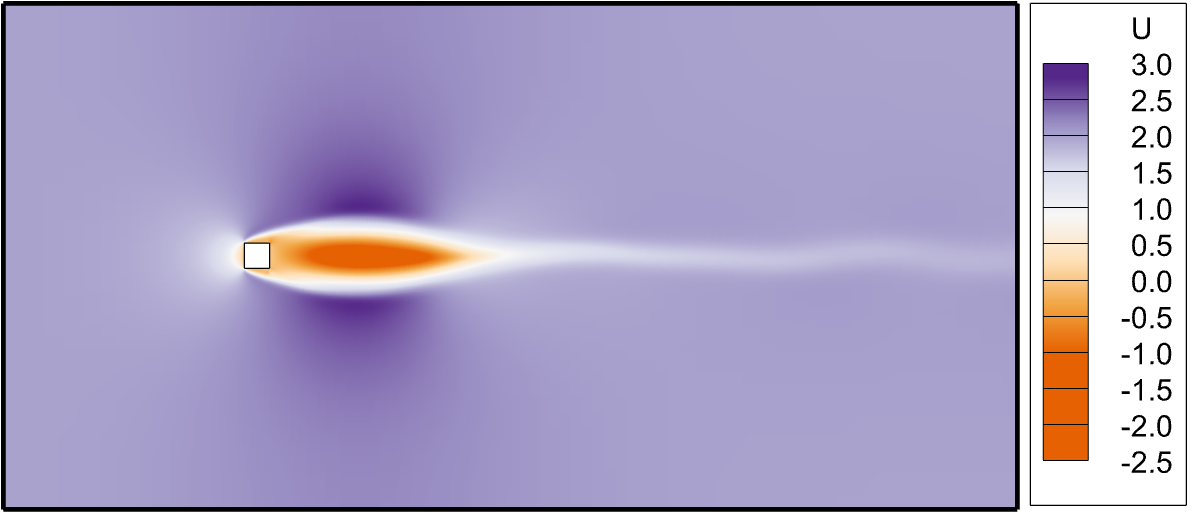}
    \caption{$Re=200$, $t_2$}
    \label{fig:1106Re200}
    \end{subfigure}

    \begin{subfigure}{0.3\textwidth}
    \includegraphics[width=\textwidth]{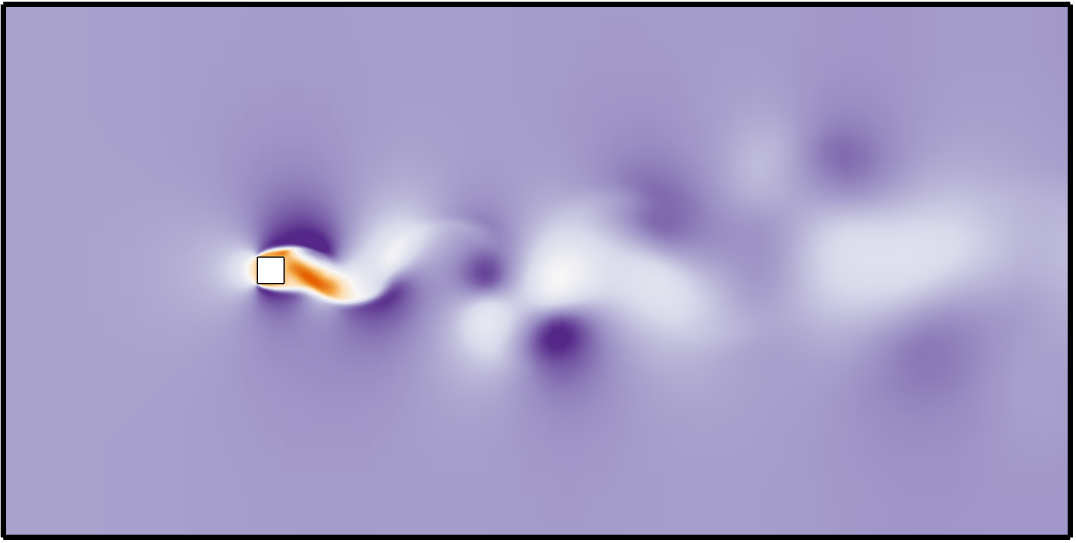}
    \caption{$Re=300$, $t_0$}
    \label{fig:1107Re300}
    \end{subfigure}
    \begin{subfigure}{0.3\textwidth}
    \includegraphics[width=\textwidth]{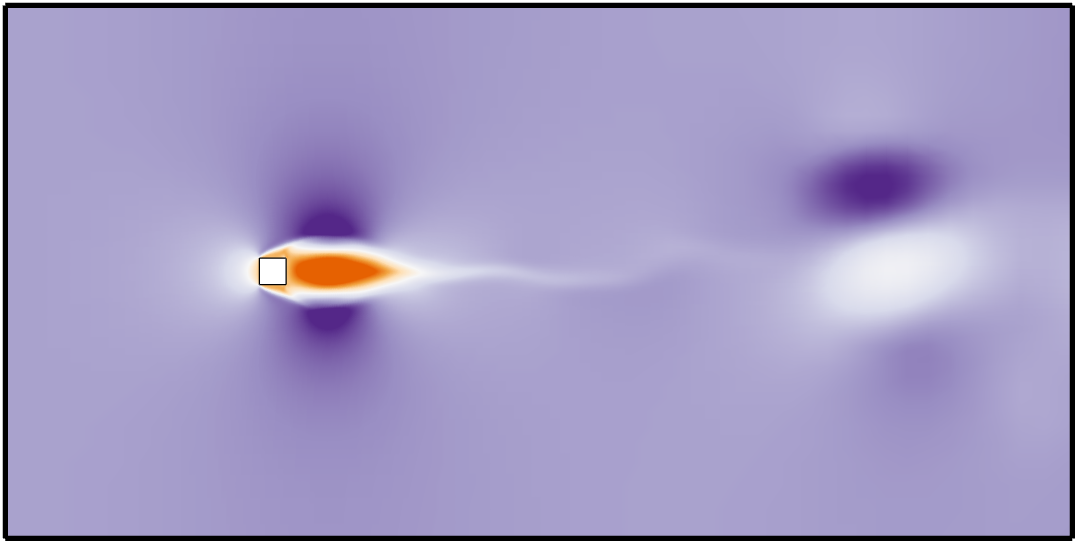}
    \caption{$Re=300$, $t_1$}
    \label{fig:1108Re300}
    \end{subfigure}
    \begin{subfigure}{0.35\textwidth}
    \includegraphics[width=\textwidth]{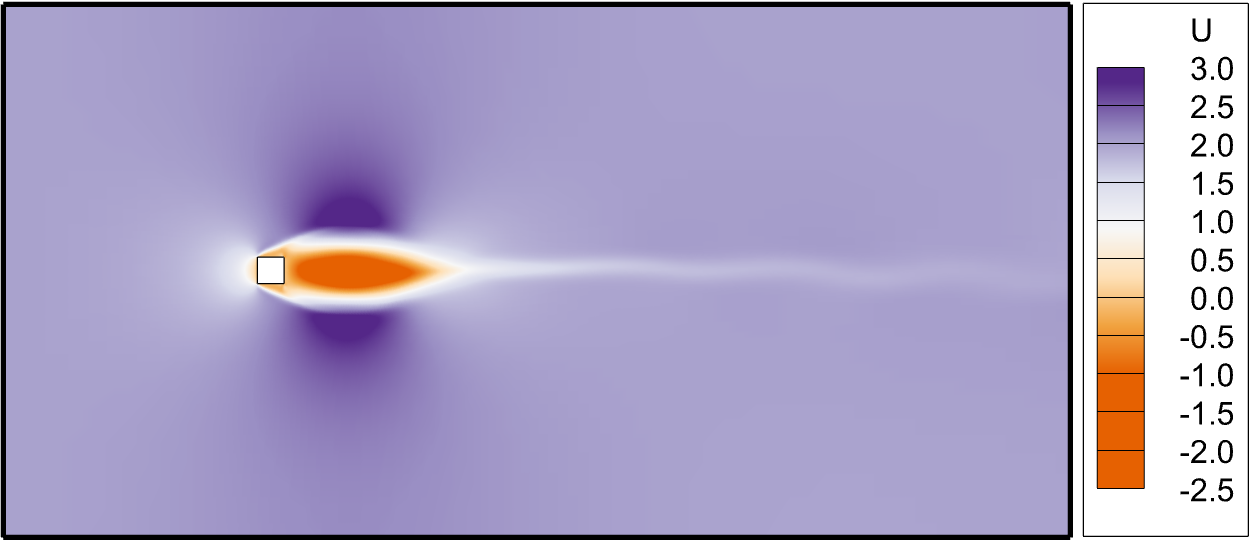}
    \caption{$Re=300$, $t_2$}
    \label{fig:1109Re300}
    \end{subfigure}

    \begin{subfigure}{0.3\textwidth}
    \includegraphics[width=\textwidth]{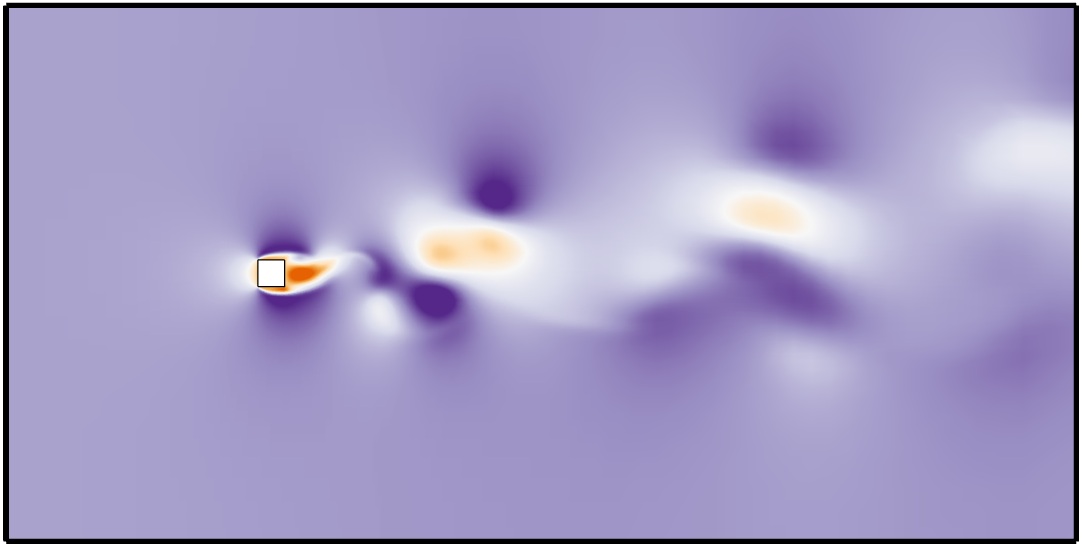}
    \caption{$Re=400$, $t_0$}
    \label{fig:1110Re400}
    \end{subfigure}
    \begin{subfigure}{0.3\textwidth}
    \includegraphics[width=\textwidth]{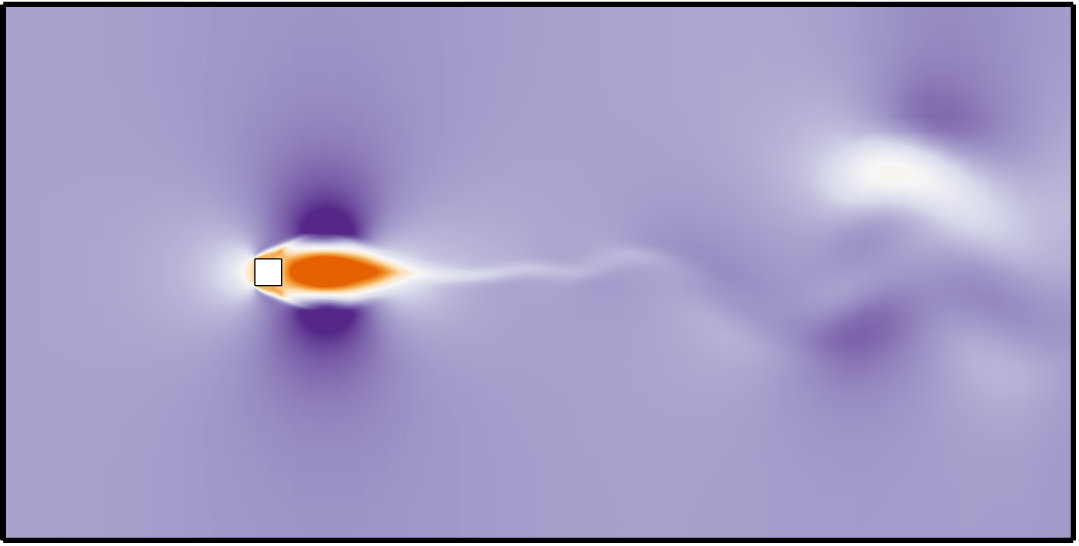}
    \caption{$Re=400$, $t_1$}
    \label{fig:1111Re400}
    \end{subfigure}
    \begin{subfigure}{0.35\textwidth}
    \includegraphics[width=\textwidth]{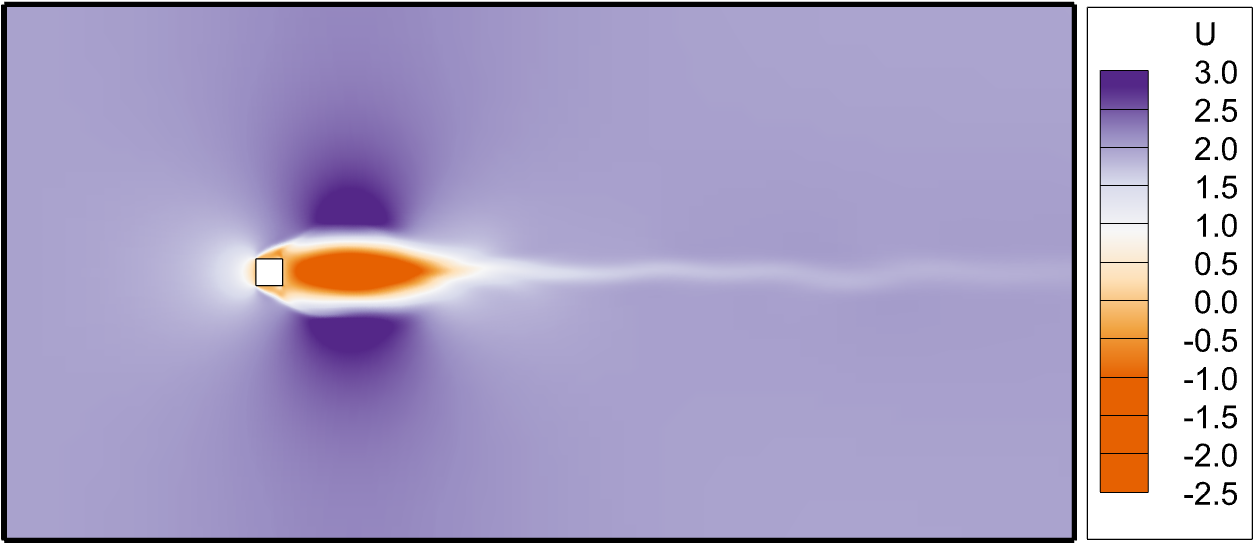}
    \caption{$Re=400$, $t_2$}
    \label{fig:1112Re400}
    \end{subfigure}
    \caption{Instantaneous velocity contours for a range of Reynolds numbers (\(Re = 100\), 200, 300, and 400) captured at successive temporal instances. Here, \(t_0\) denotes the initial time, illustrating the uncontrolled velocity field. \(t_1\) corresponds to the controlled state during the implementation of the DRL-based flow control algorithm. \(t_2\) conveys the velocity field following the cessation of control, thereby reflecting the results of the DRL algorithm's application. (a) $Re =100$, at $t_0$. (b) $Re =100$, at $t_1$. (c) $Re =100$, at $t_1$. (d) $Re =200$, at $t_0$. (e) $Re =200$, at $t_1$. (f) $Re =200$, at $t_1$. (g) $Re =300$, at $t_0$. (h) $Re =300$, at $t_1$. (i) $Re =300$, at $t_1$. (j) $Re =400$, at $t_0$. (k) $Re =400$, at $t_1$. (l) $Re =400$, at $t_1$.}
    \label{fig:velocity_contour}
\end{figure*}

As reflected by various statistical measures, the DRL agent is capable of finding optimal control strategies for reducing the overall drag. To elucidate the underlying control mechanism, a meticulous examination of the flow field is imperative. Here, we aim to decode the interaction dynamics between the synthetic jet actuation and the surrounding flow to facilitate an understanding of the DRL algorithm in AFC problems.

\cref{fig:velocity_contour} illustrates the velocity contours at three pivotal moments during the flow control process around the square cylinder for four different $Re$. \(t_0\) marks the commencement of control, representing the velocity field of the uncontrolled flow. \(t_1\) corresponds to an intermediate moment during the control process, and \(t_2\) displays velocity field of the controlled flow. This sequential representation serves to visually capture the evolution of the flow field from its initial state, through the application of control strategies, to the final controlled state, thereby providing insights into the effectiveness and impact of the control measures implemented. 

Starting at $Re = 100$, \cref{fig:1101Re100} displays the wake flow around the square cylinder without any flow control interventions. The flow regime around the square cylinder is characterized by significant flow separation around the body and a von K\'{a}rm\'{a}n vortex street within the wake region. The unsteady vortex shedding is periodic with a single shedding frequency. When jet control is introduced (as seen in \cref{fig:1102Re100}), by simultaneous blowing and suction, it substantially elongates the separation zone behind the square cylinder. As a result, the onset of vortex shedding is delayed to a position further downstream than observed in the uncontrolled case. By \cref{fig:1103Re100}, the jet control has fully influenced the wake flow, and the length of the separation zone has developed to its maximum extent. There is minimal vortex shedding observed in the far downstream of the wake flow, which is indicative of a more stable flow regime.

As we increase the $Re$, the flow fields in \cref{fig:1104Re200,fig:1107Re300,fig:1110Re400} demonstrate a transition towards increasingly unstable and complex vortex shedding behavior. The extent of fluctuation amplifies significantly. Specifically, at $Re$ of 300 and 400, the base flow (at $t_0$) is characterized by chaotic vortex shedding with strong asymmetry. This can be quantified by the multi-frequency PSD analysis in \cref{fig:baseline1}. Nonetheless, the introduction of synthetic jets influences the wake in a similar fashion across all examined $Re$ at $t_1$, as shown in \cref{fig:1105Re200,fig:1108Re300,fig:1111Re400}. In each instance, recirculation zones emerge behind the square body, which pushed the vortical structures previously shed downstream. In \cref{fig:1106Re200,fig:1109Re300,fig:1112Re400} where the recirculation zones are given time to stabilize, the von K\'{a}rm\'{a}n vortex street is absent. The flow is characterized by a marked reduction in unsteady behavior, signifying the efficacious stabilization through the deployment of DRL-based flow control strategies.

We would like to highlight that, to the best of our knowledge, the successful suppression of vortex shedding demonstrated in this wide range of $Re$ has not been observed in existing literature. At $Re=100$, 
\citeauthor{wangDRLinFluids} and \citeauthor{chen2023deep} showed successful drag reduction through DRL-based AFC control, but did not achieve the desired effect of suppressing vortex shedding in most cases\cite{wangDRLinFluids,chen2023deep}.
Notably, in the work by \citeauthor{chen2023deep}, only when the jet actuator was positioned at the rear corners of the square cylinder could the vortices in the wake field of the square cylinder be stabilized, making it the only test case to date capable of completely controlling vortex shedding in the square cylinder wake.
Other related studies include those by \citeauthor{yan2023stabilizing}\cite{yan2023stabilizing} at \(Re=500\), \(1000\), and \(2000\), and \citeauthor{yan2024aero}\cite{yan2024aero} at \(Re=1000\), which demonstrate promising results in drag and lift control around the square cylinder. However, vortex shedding still persists in the wake region under control. In our work, we have showcased the robustness of DRL-based frameworks in devising effective flow control With successful elimination of vortex shedding. A flow around square cylinders with stable wake not only gives many desirable engineering features such as greater drag reduction and minimal lateral vibration, but also paves a way for us
to study its control mechanism.

\begin{figure*}[ht]
    \centering
    \begin{subfigure}{0.48\textwidth}
    \includegraphics[width=\textwidth]{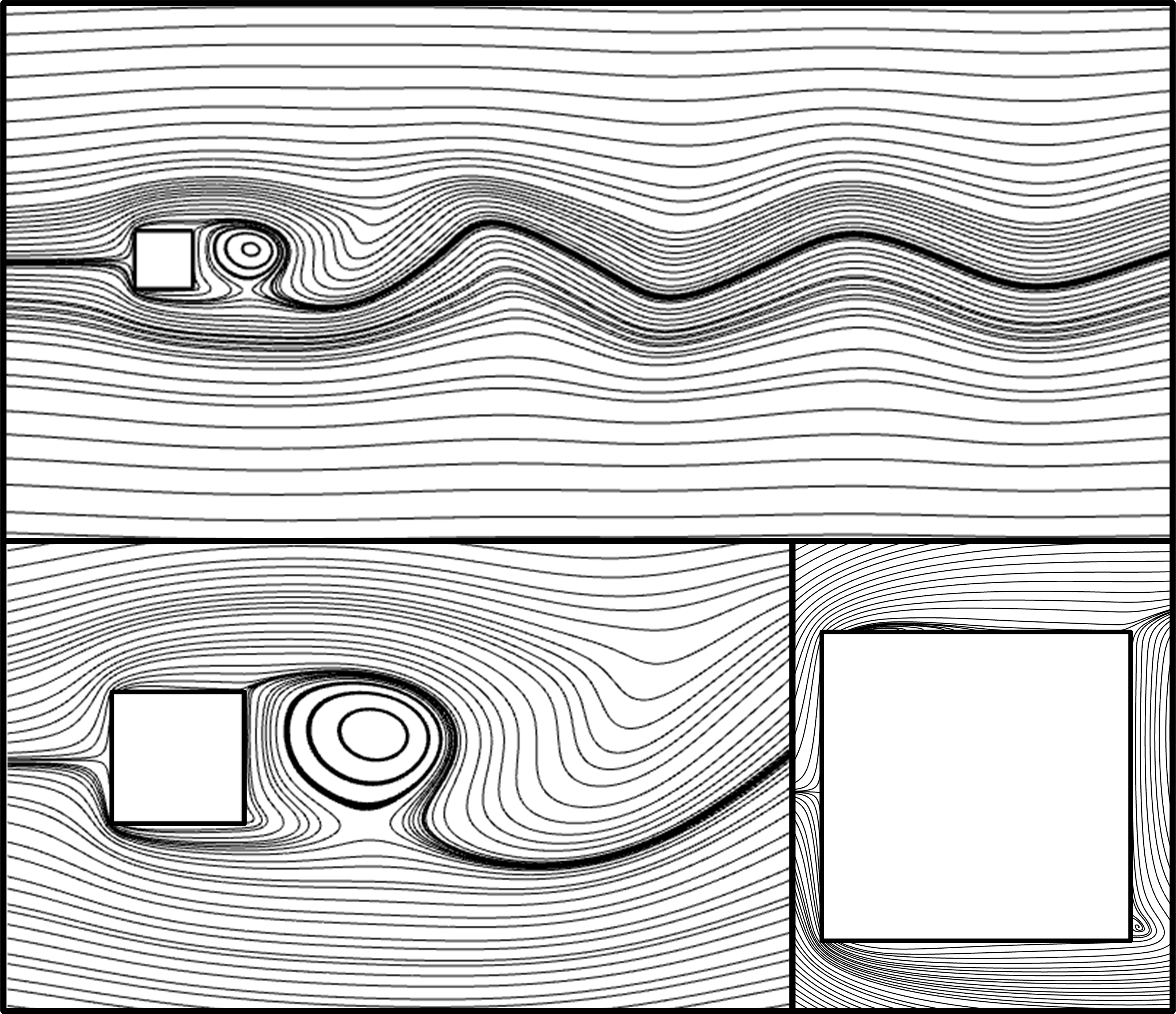}
    \caption{$Re=100$, Baseline}
    \label{fig:streamline1201}
    \end{subfigure}
    \begin{subfigure}{0.48\textwidth}
    \includegraphics[width=\textwidth]{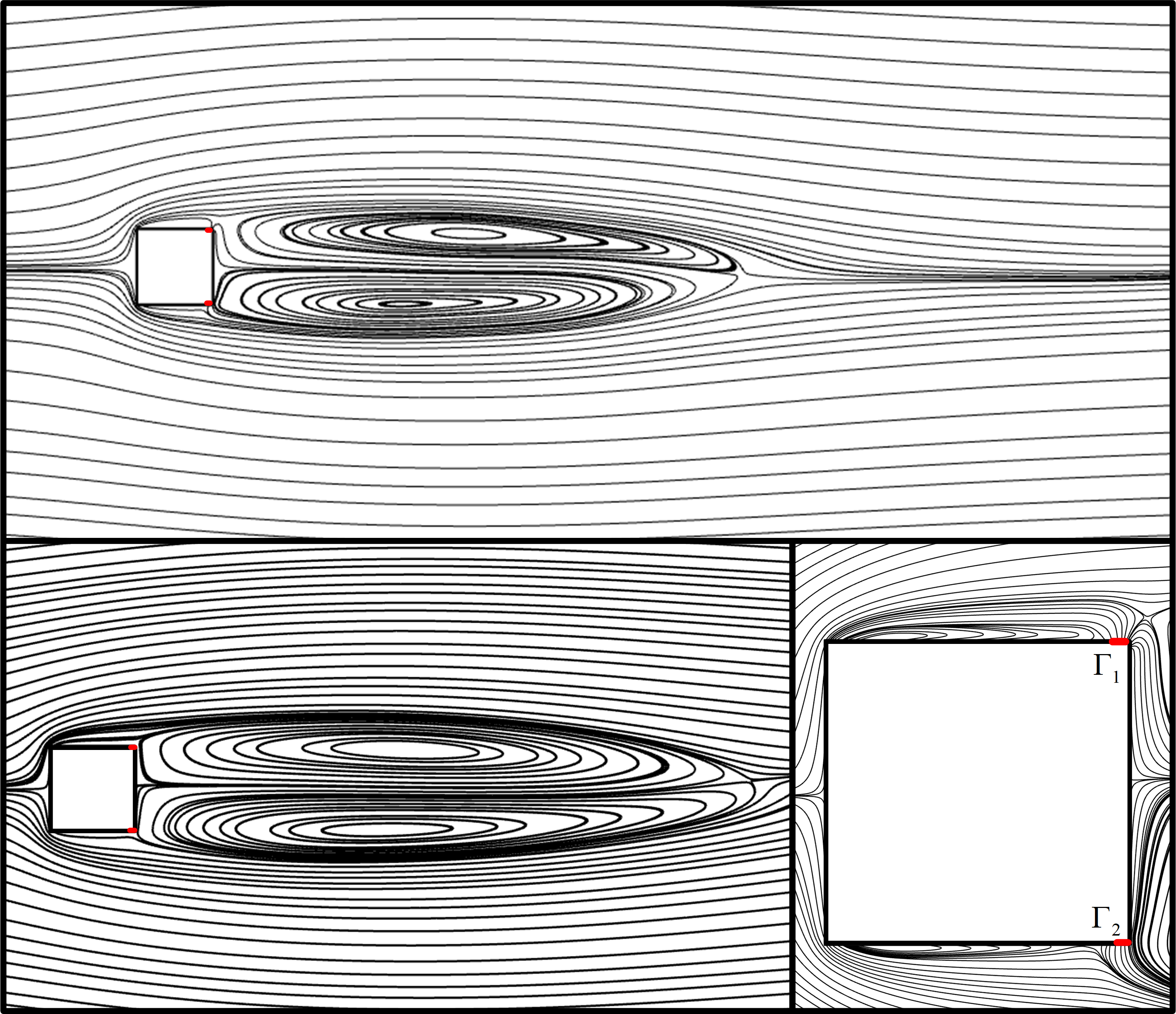}
    \caption{$Re=100$, Controlled}
    \label{fig:streamline1202}
    \end{subfigure}
    \caption{Streamline comparison of uncontrolled flow (Baseline flow) and controlled flow at $Re=100$. The upper part of the figure is the wake field of the square cylinder, and the lower part is a detailed view of the flow field around the square cylinder, showing the streamlines downstream of and around the square cylinder.}
    \label{fig:streamline12}
\end{figure*}

\begin{figure*}[htbp]
    \centering
    \begin{subfigure}{0.48\textwidth}
    \includegraphics[width=\textwidth]{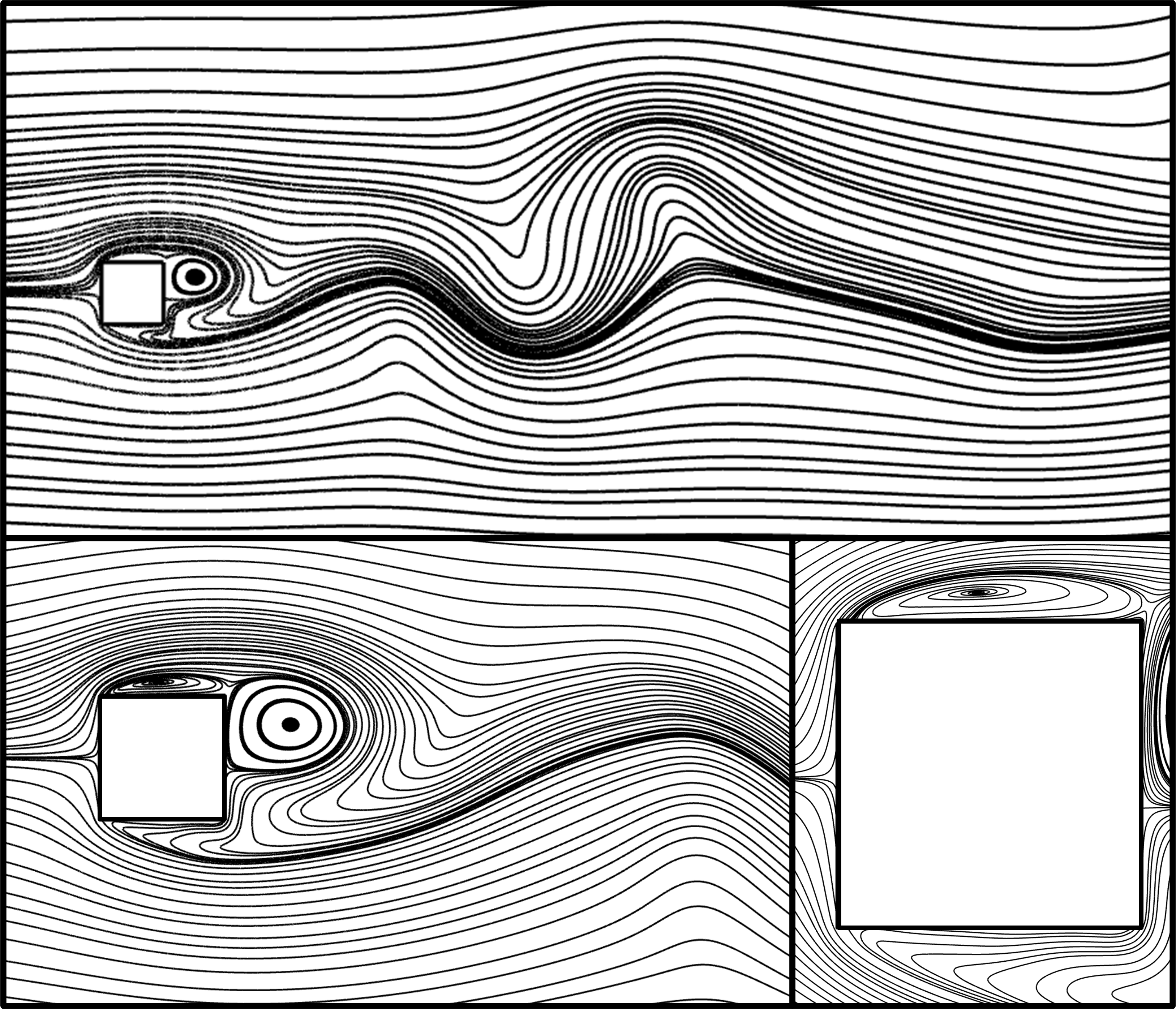}
     \caption{$Re = 400$, Baseline}
    \label{fig:streamline1301}
    \end{subfigure}
    \begin{subfigure}{0.48\textwidth}
    \includegraphics[width=\textwidth]{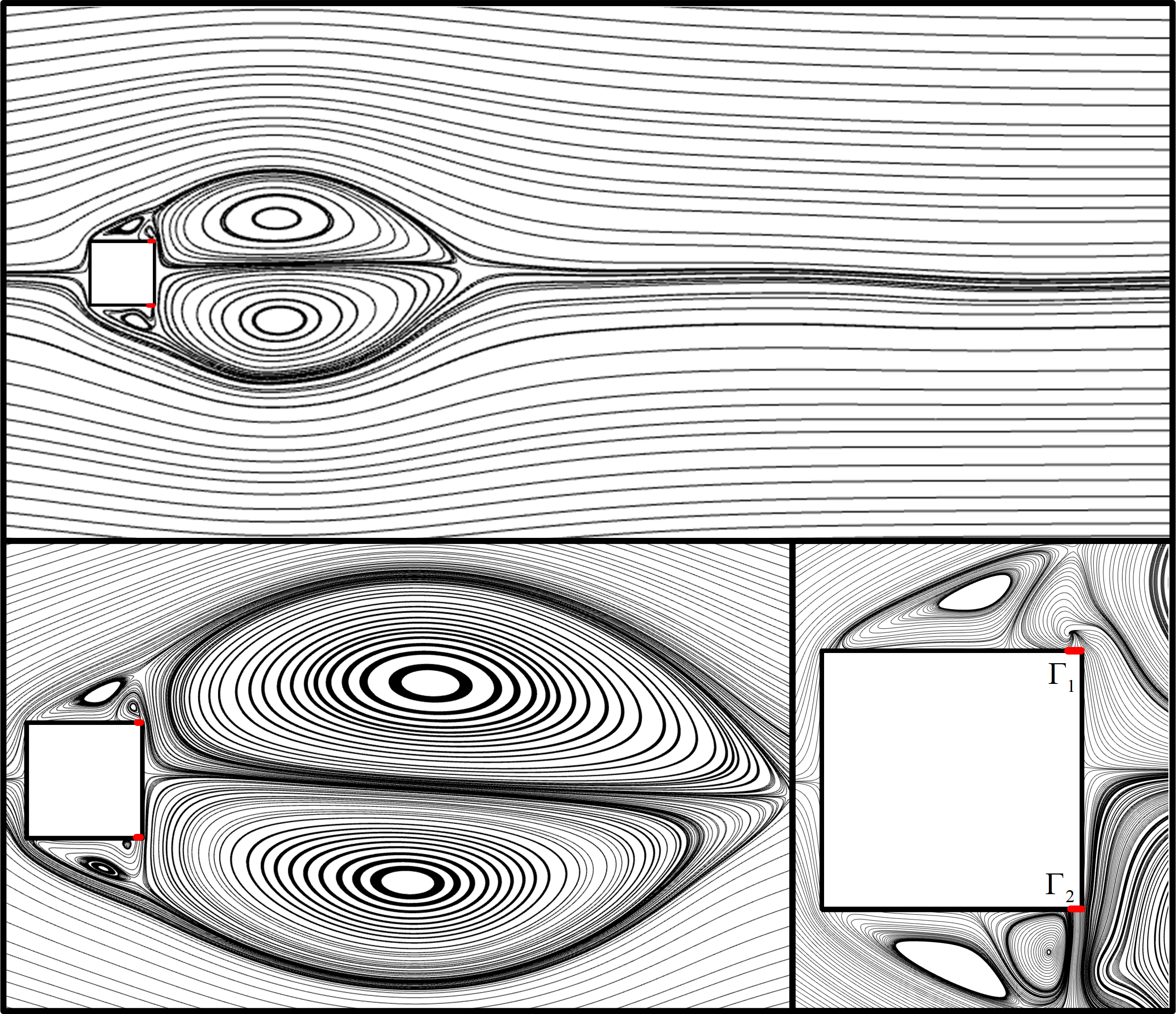}
    \caption{$Re = 400$, Controlled}
    \label{fig:streamline1302}
    \end{subfigure}
    \label{fig:streamlineRe400}
    \caption{Streamline comparison of uncontrolled flow (Baseline flow) and controlled flow at $Re = 400$.}
\end{figure*}

To understand the underlying mechanism of flow control provided by the DRL agent, we investigate the instantaneousstreamlines at $Re = 100$ and $Re = 400$ for the uncontrolled baseline and the controlled state. In \cref{fig:streamline1201}, we observe the formation of a single vortical structure shedding from the square cylinder. A detailed examination in the vicinity of the body shows that, as flow approaches the square body, it separates at the two corners, reattaches along the two lateral sides, and ultimately separates again, which results in the formation of the vortex at the rear of the body. In the absence of control, the vortex shedding process is inherently unsteady and alternates between the two sides, leading to the formation of a vortex street. In the controlled state, \cref{fig:streamline1202} shows the emergence of an elongated recirculation zone with a pair of symmetric F\"{o}ppl vortices \cite{foppl1913wirbelbewegung} in equilibrium. With the zoom-in view, it is evident that both separation bubbles on the two lateral sides of the square cylinder are still present, but the strategic placement of two jets towards the cylinder's rear significantly alters the flow dynamics. At this particular instant, the suction at $\Gamma_1$ and blowing at $\Gamma_2$ provides a more symmetrical flow pattern around the two rear corners. This adjustment fosters the development and stabilization of a recirculation zone, highlighting the DRL agent's capacity to manipulate flow characteristics through precise interventions.

At $Re = 400$, the intervention due to jet actuation is more evident. \cref{fig:streamline1301} shows the vortex shedding streamline pattern. A significantly thicker separation bubble can be observed on one side of the square due to the increased $Re$. 
The implementation of active flow control yields a recirculation zone as depicted in \cref{fig:streamline1302}, which, although shorter and thicker compared to that observed in \cref{fig:streamline1202}, still produces nearly horizontal streamlines in the wake. This indicates that the F\"{o}ppl vortices have achieved a state of equilibrium.
However, a zoom-in view reveals a different control mechanism at play. Instead of directly interrupting the separation bubbles on the sides, the two synthetic jets generate small vortical structures that act as barriers, preempting the upstream propagation of separation bubbles towards the sharp rear corners of the square cylinder. This approach also facilitates the concurrent development of separation bubbles on both sides of the cylinder. Such symmetry is instrumental in the stable emergence of F\"{o}ppl vortices directly behind the square cylinder, effectively precluding the formation of vortex shedding.

It is noteworthy that no explicit information pertaining to the control mechanism was provided throughout the training phase of the SAC agent. The only constraint was the limitation on the magnitude of the jet flow for energy efficiency.
Nevertheless, the DRL algorithm demonstrated remarkable proficiency in identifying an optimal control strategy, which is tailored to the specific characteristics of the flow.
This adaptability underscores the robustness of the DRL framework in addressing AFC problems. Such autonomy is particularly valuable in scenarios where the flow dynamics are complex and highly variable, as it allows the algorithm to adapt its control strategy to optimize performance across a broad spectrum of conditions.

\section{Conclusions}\label{sec:Conclusions} 

The current study methodically investigates the application of DRL-based AFC through the actuation of synthetic jets on a confined squared cylinder.
We leverage the SAC algorithm to precisely control the mass flow rate of synthetic jets located on both the upper and lower sides of a squared cylinder. By altering the formation and shedding of vortices as the fluid navigates past the squared cylinder, the control strategy informed by DRL successfully mitigates lift and reduces drag.

\begin{itemize}
    \item Initially, computational analysis was performed on the flow around the confined squared cylinder at $Re = 100$, 200, 300, and 400, encompassing both transitional phases and the establishment of stable periodic vortex shedding stages. Under the flow conditions at $Re = 100$ and 200, the wake behind the squared cylinder evolved through a transitional phase into a state of stable periodic vortex shedding. Spectral analysis results indicated that at these two $Re$, the vortex shedding frequency associated with the $C_L$ displayed characteristics of mono-frequency. 
    In contrast, the flow configurations at $Re = 300$ and $Re = 400$ resulted in irregular and unstable vortex shedding behind the square cylinder. Spectral analysis revealed that under these flow conditions, the vortex shedding frequency was dominated by multiple frequencies, diverging from the singular frequency shedding observed at lower $Re$.
    The dominance of multiple frequencies in vortex shedding around the squared cylinder implies a more complex flow pattern downstream of the squared cylinder, thereby escalating the challenge of implementing flow control through DRL algorithms.
    \item We utilizes the SAC algorithm for active flow control with synthetic jets on a square cylinder. Results show that DRL agents exhibit proficiency in acquiring effective control methods across a range of $Re$, with higher $Re$ presenting greater challenges and variability in reward trajectories. The SAC-based flow control results show that under four different $Re$ flow configurations, the $C_L$ quickly drops to a level close to zero after the control is initiated and stabilizes at a level close to zero. Simultaneously, the shedding frequency of vortices decreases, eliminating the prominent peaks observed in spectral analysis under baseline flow conditions.
    The $C_D$ undergo significant oscillations during synthetic jet activation before quickly decreasing and stabilizing near their minimum values. The previously observed dominant single- or multi-frequency modes within the baseline flow are disrupted and eliminated, indicating the effective removal of regular or chaotic vortex shedding phenomena by the synthetic jet. Quantification of average $C_D$ demonstrates a significant reduction at $Re$ of 100, 200, 300, and 400 by approximately 14.4\%, 26.4\%, 38.9\%, and 47.0\%, respectively. 
    With the aim of reducing resistance and suppressing lift, we hope that the mass flow rate of the jet used can remain at a relatively low level to ensure that the control strategy meets energy-saving requirements. The strategy provided by SAC has met our expectations in this regard.
 
    \item An in-depth analysis of the flow field pre and post control is conducted to enhance understanding of the agent's control strategy. From the velocity contour plots under both uncontrolled and controlled flow conditions at $Re = 100$, 200, 300, and 400, it is observed that in the wake of the uncontrolled flow around the squared cylinder at $Re = 100$ and 200, there is a regular, alternating vortex shedding phenomenon. However, at $Re = 300$ and 400, the vortex shedding as the fluid flows past the squared cylinder becomes irregular and asymmetric. After the implementation of AFC, the periodic shedding of vortices in the wake of the square cylinder is effectively suppressed. Importantly, the irregular and unsteady vortex shedding phenomena are also successfully mitigated. The streamline contour maps clearly demonstrate that under the influence of the synthetic jets, the recirculation region downstream of the square cylinder is significantly elongated and expanded. The establishment of F\"{o}ppl vortices in equilibrium indicate that the resultant flow is stable and steady. A close examination of the flow around the square body reveals different control mechanisms at varying $Re$. This demonstrates the robustness of the DRL-based AFC technique in devising catered control strategies for distinct flow field to adapt for optimal performance.

\end{itemize}

In conclusion, across various $Re$, the deployment of intelligent agents trained via the SAC algorithm has facilitated the learning of a control strategy capable of dynamically adjusting the mass flow rate of synthetic jets. This strategy significantly diminishes lift around the squared cylinder, substantially reduces drag, and suppresses the generation and shedding of vortices within the complex, multi-frequency dominated wake. These research outcomes underscore the adeptness of DRL in controlling intricate, multi-frequency dominated vortex shedding phenomena, and highlight the robustness of $Re$ to AFC techniques surrounding a confined squared cylinder based on DRL.

\begin{acknowledgments}

The authors thank Dr. Jean Rabault (University of Oslo, Oslo, Norway) and Mr. Qiulei Wang (The University of Hong Kong, Hong Kong SAR, China) for making their open-source codes for deep reinforcement learning and numerical simulation available online at \href{https://github.com/jerabaul29/Cylinder2DFlowControlDRLParallel}{https://github.com/jerabaul29/Cylinder2DFlowControlDRLParallel} \cite{rabault2024cylinder2dflowcontroldrlparallel}, and \href{https://github.com/venturi123/DRLinFluids}{https://github.com/venturi123/DRLinFluids} \cite{DRLinFluids}.
\end{acknowledgments}

\section*{Declaration of interests}
The authors report no conflict of interest.


\section*{Author ORCIDs}

\noindent Wang Jia \href{https://orcid.org/0009-0008-2786-397X}{https://orcid.org/0009-0008-2786-397X}\\
Hang Xu \href{https://orcid.org/0000-0003-4176-0738}{https://orcid.org/0000-0003-4176-0738}.

\section*{References}
\nocite{*}
\bibliography{aipsamp}
\end{document}